\documentclass[aps,prb, reprint,twocolumn,superscriptaddress,showpacsx]{revtex4-1}

\usepackage{amsmath,amssymb}
\usepackage{graphicx}
\usepackage{xcolor}
\usepackage[colorlinks]{hyperref}
\hypersetup{
  linkcolor  = blue,
  citecolor  = blue,
  urlcolor   = blue,
  colorlinks = blue,
}

\usepackage[final]{changes}

\definechangesauthor[color=red]{AP}
\definechangesauthor[color=blue]{GK}

\begin{document}

\title{Fermi-Surface Selective Determination of the $\mathbf{g}$-Factor Anisotropy in URu$_2$Si$_2$}

\author{Ga\"{e}l Bastien}
\altaffiliation{Present address: Leibniz-Institute for Solid State Research  (IFW) Dresden, Helmholtzstr.~20, 01069 Dresden, Germany.}
\affiliation{Univ.~Grenoble Alpes, CEA, IRIG-PHELIQS, F-38000 Grenoble, France}
\author{Dai Aoki}
\affiliation{Univ.~Grenoble Alpes, CEA, IRIG-PHELIQS, F-38000 Grenoble, France}
\affiliation{IMR, Tohoku University, Oarai, Ibaraki 311-1313, Japan}
\author{G\'erard Lapertot}
\affiliation{Univ.~Grenoble Alpes, CEA, IRIG-PHELIQS, F-38000 Grenoble, France}
\author{Jean-Pascal Brison}
\affiliation{Univ.~Grenoble Alpes, CEA, IRIG-PHELIQS, F-38000 Grenoble, France}
\author{Jacques Flouquet}
\affiliation{Univ.~Grenoble Alpes, CEA, IRIG-PHELIQS, F-38000 Grenoble, France}
\author{Georg Knebel}
\email{georg.knebel@cea.fr}
\affiliation{Univ.~Grenoble Alpes, CEA, IRIG-PHELIQS, F-38000 Grenoble, France}

\date{\today } 

\begin{abstract}

The $g$-factor anisotropy of \added{the heavy quasiparticles} in the hidden order state of URu$_2$Si$_2$ has been determined from the superconducting upper critical field and microscopically from Shubnikov-de Haas (SdH) oscillations. We present a detailed analysis of the $g$-factor for the $\alpha$, $\beta$ and $\gamma$ Fermi-surface pockets. Our results suggest a strong $g$-factor anisotropy between the $c$ axis and the basal plane \added{for all observed Fermi surface pockets}. \deleted{For the $\beta$ and $\gamma$ Fermi surface pockets we find that the $g$-factor varies between $g \approx 2.5$ for field along the $c$ axis, and a vanishing $g$ for field applied in the basal plane of the tetragonal crystal. In addition,  we observed an unexpected $g$-factor anisotropy in the basal plane for the $\alpha$ branch with non-vanishing $g$ for field applied along [110].The $\beta$-branch shows a nonlinear field dependence of the quantum oscillation frequency, leading to a reduction of the observed $g$-factor anisotropy under magnetic field.   The  comparison with the anisotropy of the $g$-factor determined from the upper critical field confirms the importance of the heavy branches for the superconducting pairing.} \added{The observed anisotropy of the $g$-factor from the quantum oscillations is in good agreement with the anisotropy of the superconducting upper critical field at low temperatures, which is strongly limited by the paramagnetic pair breaking along the easy magnetization axis $c$. \added{However, the anisotropy of the initial slope of the upper critical field near $T_c$ cannot be explained by the anisotropy of the effective masses and Fermi velocities derived from quantum oscillations.}}

\end{abstract}


\maketitle

\section{Introduction}
 
 The "hidden order" state in the heavy-fermion compound URu$_2$Si$_2$ that develops below $T_0 = 17.5$~K is still under debate despite several decades of research after its discovery.\cite{Palstra1985} Intense experimental effort has been employed, but right now no spectroscopic probe could unambiguously identify the order parameter\deleted{ or the excitation modes at the ordering wave vector}. A wide variety of order parameter scenarios have been proposed, most of them based on higher multipolar ordering, various kinds of density wave ordering, or hybridization of the $5f$-states with the conduction electrons as order parameter itself. Recent reviews on the theoretical and experimental status are given in Refs.~\onlinecite{Mydosh2011, Mydosh2014a, Mydosh2014}. Novel proposals \deleted{ based on Raman spectroscopy results 
 is } are a chirality-density wave groundstate of hexadecapoles,\cite{Kung2015} \added{or odd-parity electric dotriacontapolar order.\cite{Kambe2018} }

In addition to the hidden order state, an unconventional superconducting state is formed below $T_{sc}=1.5$~K, which coexists with the hidden order. This superconducting state is characterized by spin singlet pairing.\cite{Hattori2018} Recent thermal conductivity and specific heat measurements support a chiral $d$-wave superconducting gap structure characterized by horizontal line nodes and point nodes at the poles.\cite{Kasahara2007, Kittaka2016} The spontaneous breaking of time-reversal symmetry is in accordance with the experimentally detected chiral $d$-wave state.\cite{Li2013, Yamashita2014, Kawasaki2014, Schemm2015} 

Like in many heavy-fermion systems, the magnetic susceptibility in URu$_2$Si$_2$ shows at high temperatures a Curie-Weiss behavior indicating a local moment behavior. Below 70~K, hybridization between the 5$f$ states and the $spd$ electrons of the ligands sets in, and heavy quasiparticle bands are formed.\cite{Haule2009, Bachar2016} At low temperatures, in the heavy-fermion state, the static bulk susceptibility as well as the dynamical spin susceptibility show a large anisotropy between the $c$-axis and the $a$-axis of the tetragonal crystal.\cite{Palstra1985, Kohori1996, Emi2015, Hattori2016}  Magnetic excitations detected by neutron scattering are strictly longitudinal, indicating Ising-type magnetic fluctuations in URu$_2$Si$_2$.\cite{Broholm1991, Bourdarot2010} Measurements of the non-linear magnetic susceptibility confirm this Ising character of the magnetic response.\cite{Trinh2016}

The strong Ising character of the $5f$ electrons in URu$_2$Si$_2$ has been also confirmed on the basis of density functional theory (DFT) electronic structure calculations.\cite{Werwiniski2014} The \deleted{strong} Ising anisotropy arises from a combination of the peculiar Fermi surface nesting and strong spin-orbit interaction. While in this electronic structure calculations the $5f$ electrons are treated fully itinerant, other models supposing a localized 5$f^2$ non-Kramers doublet ground state could explain the large magnetic uniaxial anisotropy as well.\cite{Chandra2013, Chandra2015} In the localized picture the Ising character of the localized $f$ states of the uranium ions is transferred by hybridization to the  heavy quasiparticles forming a Fermi surface.  However,  the magnetic and crystal electric field  ground-state wave function in URu$_2$Si$_2$ is still under discussion \cite{Sundermann2016} and even the localized or itinerant character of the 5$f$ electrons\deleted{remains an open question}. 

\added{In the present article, we study in detail the $g$-factor anisotropy \added{for three different Fermi surface pockets} in this tetragonal system. The orientation of the sample was tuned to study field directions between [001] and [100] and between [001] and [110], as well as within the basal plane for the quantum oscillation and the upper critical field measurements. So we could determine the $g$-factor anisotropy for \deleted{each}\added{different} Fermi surface pockets in this multiband system. We compare the Fermi-surface selective $g$-factor \added{from the quantum oscillations} to  the effective $g$-factor determined from the anisotropy of the upper critical field. \deleted{Differences in the $g$-factor anisotropy can be traced by different $j_z$ character of the $5f$ electrons and also to different hybridization with the other $s, p, d$ electrons. } }

\added{
The $g$-factors determined for each Fermi surface show an anisotropy between the $c$-axis and the basal plane. 
\deleted{For the $\alpha$ branch we find an additional anisotropy of the $g$-factor for fields applied along  [100] and [110].} 
\added{In addition we show that the observed $g$-factor of branch $\beta$ is field dependent.  The analysis of the spin slitting zero of the $\alpha$ branch is more delicate, as we observe 17 spin-splitting zero in the (010) plane and only 12 zeros in the (110) plane. This implies either a non-monotonously varying $g$-factor in the (010) plane, or the observation of spin-splitting zeros in the basal plane, which could not be resolved in the present experiment. } 
\deleted{However, as the $g$-factor from quantum oscillations can only be determined up to an additional constant, this anisotropy is less important if $g$ is rather high (although it is unlikely) and for all field directions $g_\alpha \gtrsim 3$.} Our results strongly suggest that the Fermi surface pockets with strongly anisotropic $g$-factor determine the superconducting upper critical field at low temperature. The superconducting pairing  is known (from the large specific heat jump or the high orbital limitation) to be dominantly governed by the heaviest quasiparticle pockets \added{with an strongly anisotropic Fermi velocities}. \deleted{The light hole Fermi surface is not the main Fermi surface for the superconducting pairing. Comparing the anisotropy of the initial slope to the Fermi velocities of the observed Fermi surface pockets, we conclude that the heaviest with strongly anisotropic Fermi velocity has still not be observed in quantum oscillations in URu$_2$Si$_2$.}
In the present state of band-structure calculations in heavy-fermion compounds, there is no derivation of the $g$-factor right at the Fermi level  nor of its link with the bulk Pauli susceptibility. The interest of this study is to give an experimental framework for future theoretical developments. 
}

\subsection{Fermi surface of URu$_2$Si$_2$}

The Fermi surface of URu$_2$Si$_2$ has been previously studied in detail by magnetic quantum oscillations,\cite{Bergemann1997, Ohkuni1999, Shishido2009, Hassinger2010, Altarawneh2011, Aoki2012, Scheerer2014} cyclotron resonance measurements,\cite{Tonegawa2012, Tonegawa2013}, and  angular resolved photoemission spectroscopy (ARPES).\cite{Santander-Syro2009, Yoshida2010, Dakovski2011, Yoshida2013, Meng2013, Chatterjee2013, Bareille2014, Durakiewicz2014} According to these experiments and to their comparison with band-structure calculations,\cite{Elgazzar2009, Oppeneer2010, Ikeda2012} four different Fermi surface sheets have been observed. At the center of the simple tetragonal Brillouin zone, a rather isotropic large hole Fermi surface $\alpha$ exists. The electron Fermi surface $\beta$ is four-folded and located between the $\Gamma$ and $X$ points of the simple tetragonal Brillouin zone. A small elliptical electron Fermi surface $\gamma$ and a  heavy Fermi surface $\eta$ are located either at the $M$ point or at the $\Gamma$ point. As URu$_2$Si$_2$ is a compensated metal, we can conclude that the quantum oscillation experiments failed up to now to detect a heavy electron pocket which, following band structure calculations,\cite{Elgazzar2009, Ikeda2012} is located at the $M$ point of the Brillouin zone. A four-armed cage-like Fermi surface around the $\alpha$ pocket is expected in Ref.~\onlinecite{Ikeda2012}, while it disappears completely in other calculations.\cite{Elgazzar2009} No  orbit corresponding to this cage-like structure has been detected in quantum oscillation experiments. Thus, the Fermi surface of URu$_2$Si$_2$ has not been completely determined and it is not fully understood.  

\subsection{Detection of the $g$-factor}

The Ising-type characteristics of the quasiparticles forming the Fermi surface \added{in URu$_2$Si$_2$} has been supported from the analysis \cite{Altarawneh2012} of old quantum oscillation experiments.\cite{Ohkuni1999} 
This has been inferred from the observation of so-called spin-splitting zeros in the amplitude of the quantum oscillations. 
In general, the spin splitting of the Fermi surface under magnetic field gives rise to interference of quantum oscillations from spin-up and spin-down electrons leading to a modulation of the amplitude of the quantum oscillations.  The angular dependence of the amplitude of the first harmonic is given by the spin-damping factor
\begin{equation}
a(\theta)=a_0(\theta)\left|cos(\pi g(\theta) m^\star(\theta)/2m_0)\right| ,
\label{spinzeroang}
\end{equation}
when the quantum oscillation frequencies and effective masses $m^\star$ for the spin-up and spin-down electrons are equal. The prefactor $a_0(\theta)$ contains the other factors of the Lifshitz-Kosevich formula and is expected to vary slowly with angle.\cite{Shoenberg1984} The amplitude of the quantum oscillations vanishes  when the product of the $g$-factor  and the enhancement factor of the  effective mass $(m^\star/m_0)$ is an odd integer.  This phenomenon is called a spin-splitting zero. It  allows for  the determination of the product $m^\star g$. The effective mass $m^\star$ can be determined independently from the temperature dependence of the oscillations amplitude.
Spin-splitting zeros in quantum oscillations were observed in many systems such as copper or gold and have been used to determine the angular dependence of the $g$-factor in simple metals (see e.g.~Ref.~\onlinecite{Higgins1980}). 
It is also reported for quasi-two-dimensional metals with strongly anisotropic effective masses,\cite{Bergemann2003,Wosnitza2008} or in high $T_c$ superconductors.\cite{Ramshaw2010}
However, in the case of heavy-fermion and related intermetallic compounds the observation of successive spin-splitting zeros is rather rare and has been reported only in CeIn$_3$,\cite{Settai1995} where the effective mass of the $d$-branch  is anisotropic in spite of a cubic system, and in URu$_2$Si$_2$. \deleted{ In YRu$_2$Si$_2$ only a single spin-splitting zero was detected by rotating the field angle from [001] to [100].\cite{Ikezawa1995}} 

\added{In URu$_2$Si$_2$ the observation of the spin-splitting zero has been reported only for the $\alpha$ branch in the (010) plane.\cite{Ohkuni1999} For this branch the effective mass varies from $m^\star \approx 12 m_0$ for field along the $c$ axis to  $m^\star \approx 10 m_0$ along the $a$ axis, thus it is rather isotropic. \cite{Ohkuni1999, Hassinger2010, Aoki2012} The observation of the spin-splitting zero \added{for branch $\alpha$} in URu$_2$Si$_2$ has been interpreted as signature of an Ising-type $g$-factor with $g^{[001]} = 2.5$ along the $c$ axis and a vanishing small value $g^{[100]}\approx 0$ along the $a$ axis.\cite{Altarawneh2012}}


 The $g$-factor determined from quantum oscillations is always an \deleted{orbitally} averaged $g$-factor
\begin{equation}
g = \frac{\oint g(\mathbf{k,B})v_{\mathbf{k}}^{-1}dk}{\oint v_{\mathbf{k}}^{-1}dk }
\label{orbitalaverage}
\end{equation}
over the orbit perpendicular to the applied magnetic field.\cite{Higgins1980} \added{As it dependends on ${\mathbf k}$ and the magnetic field direction ${\mathbf B}$ is a tensor quantity.} \footnote{We will use the notation e.g.~$g^{[100]}$ or $g_a$ for the component of $g$ parallel to the field applied along the $a$ axis, $H \parallel [100]$.}   It is Fermi-surface \deleted{pocket} selective, and its relation to macroscopic properties like the spin susceptibility is not at all straightforward, especially when there is duality between the localized and itinerant character of the $5f$-electrons. To calculate the spin susceptibility, one should determine the $g$-factor for every $k$-point on all the  Fermi surface pockets and average over them. As quantum oscillations are only observed on extreme orbits, it seems only possible for almost  spherical closed Fermi surfaces, and when the complete Fermi surface can be observed in the experiment. In heavy-fermion systems this is \deleted{almost never the case} \added{rare}. 

\deleted{A similar} In a superconductor, the $g$-factor anisotropy  can be determined from the \added{paramagnetic limitation} of the superconducting upper critical field $H_{c2}$. In URu$_2$Si$_2$ $H_{c2}$ varies from 3~T along the $c$ axis to 12~T in the basal plane.\cite{Brison1995, Altarawneh2012} Along the $c$-axis at low temperatures, $H_{c2}$ is determined by the paramagnetic limiting field $\mu_0H_{c2}^P = \frac{\sqrt{2} \Delta}{\mu_B g}$ which is given by the superconducting gap $\Delta$ and the effective $g$-factor for a single band isotropic superconductor.\cite{Clogston1962} 
From the angular dependence of $H_{c2}$ at 30 mK between the $c$-axis and the basal plane, taking only the paramagnetic limit into account a strongly anisotropic $g$-factor has been determined with $g^{[001]}$ = 2.65 along the $c$ axis and $g^{[100]} \approx 0.5$ for $H \parallel a$.\cite{Altarawneh2012} Lower $g$-factor values were obtained by fitting the temperature dependence of the upper critical field along the $c$ and $a$ axis taking into account the orbital limit: $g^{[001]} =1.9$ and $g^{[100]}=0.2$.\cite{Brison1995} Thus, $H_{c2}$ along the $a$ axis is close to the pure orbital limit.
The $g$-factor determined from the superconducting critical field gives, in difference to that from quantum oscillations, an average of all electrons participating to the superconducting pairing.  It is only for a single-band isotropic superconductor that it is directly related to the spin susceptibility $\chi/\chi_0 \propto g m^\star/(2 m_0)$, where $\chi_0$ and $m_0$ are  the spin susceptibility and band mass of a free electron gas.\cite{Higgins1980}

\added{Previously, splin-splitting zeros have been observed ony for the $\alpha$ Fermi surface pocket in URu$_2$Si$_2$ at many field angles in the (010) plane.\cite{Ohkuni1999} An analysis of the effective $g$-factor from these data has been reported by Altarawneh et al. in Ref.~\onlinecite{Altarawneh2012} and its anisotropy agrees remarkably well with that found from the Pauli limit of the superconducting upper critical field. However, in this previous work, only the spin-splitting zeros of the $\alpha$ pocket  in the (010) plane has been taken into account. Here we report the observation of the spin-splitting zero for all  observed Fermi surface pockets and extend previous work also to the (110) plane. }

\deleted{ 
In the present article, we study in detail the $g$-factor anisotropy \added{for three different Fermi surface pockets} in this tetragonal system. The orientation of the sample was tuned to study field directions between [[001] and [100] and [001] and [110], as well as within the basal plane for quantum oscillations and upper critical field measurements. So we could determine the $g$-factor anisotropy for each Fermi surface pocket in this multiband system. We compare this Fermi surface selective $g$-factor  to  the effective $g$-factor determined from the anisotropy of the upper critical field. Differences in the $g$-factor anisotropy can be traced by different $j_z$ character of the $5f$ electrons and also to different hybridization with the other $s, p, d$ electrons.   }

\deleted{
The $g$-factors determined for each Fermi surface show an anisotropy between the $c$-axis and the basal plane. For the $\beta$ and the $\gamma$ Fermi-surface pockets, we find $g \approx 2.5$ for field applied along the $c$-axis and a vanishing $g$-factor in the basal plane. 
\deleted{For the $\alpha$ branch we find an additional anisotropy of the $g$-factor for fields applied along  [100] and [110].} 
\added{In addition we showed that the $g$-factor of branch $\beta$ is field dependent.  The analysis of the spin slitting zero of the $\alpha$ branch is more delicate, as we have observe 17 spin-splitting zero in the (010) plane and only 12 spin-splitting zeros in the (110) plane. This implies either a non-monotonously varying $g$-factor in the (010) plane, or the observation of spin-splitting zeros in the basal plane, which could not be resolved in the present experiment. } 
\deleted{However, as the $g$-factor from quantum oscillations can only be determined up to an additional constant, this anisotropy is less important if $g$ is rather high (although it is unlikely) and for all field directions $g_\alpha \gtrsim 3$. } Our results strongly suggest that the Fermi surface pockets with strongly anisotropic $g$-factor might be more important for the superconducting pairing which is known (from the large specific heat jump or the high orbital limitation) to be dominantly governed by the heaviest quasiparticle pockets \added{with an strongly anisotropic effective mass}. \deleted{The light hole Fermi surface is not the main Fermi surface for the superconducting pairing.}
In the present state of band-structure calculations in heavy-fermion compounds, there is no derivation of the $g$-factor right at the Fermi level and nor of its link with the bulk Pauli susceptibility. The interest of this study is to give an experimental framework for future theoretical developments. 
}

\section{Experimental Details}

URu$_2$Si$_2$ crystallizes in the body-centered tetragonal ThCr$_2$Si$_2$-type crystal structure with space group $I4/mmm$. In the hidden order phase, the symmetry is lowered and the simple tetragonal unit cell volume  below 17.5~K  is twice that of the paramagnetic state. The space group of the hidden order state is still under discussion as it depends on the symmetry of the hidden order state.\cite{Harima2010, Harima2017}  
Three different URu$_2$Si$_2$ single crystals S1, S2, and S3 were used in this study. Samples S1 and S2 have been grown and investigated at CEA Grenoble, S3 has been grown and measured at IMR Oarai. The sample S1 was cut by spark erosion from a large single crystal which has been grown by the Czochralski pulling method in a tetra-arc furnace under argon atmosphere.\cite{Aoki2010} The samples S2 and S3 were grown by the indium flux method.\cite{Baumbach2014} The residual resistivity ratio $RRR=R(300$~K$)/R(0~$K) of S1, S2, and S3 are 275, 350, and 300, respectively. Resistance measurements were performed with an electrical current along the [010] direction in top-loading dilution refrigerators from Oxford Instruments down to $T=22~$mK with maximal magnetic field of 15~T, at CEA Grenoble (S1 and S2) and at IMR Oarai (S3). Due to their irregular geometry we do not calculate the resistivity and present only the measured resistance for samples S1 and S2. The samples were rotated with respect to the magnetic field using a commercial Swedish rotator which is driven by a stepper motor. The magnetoresistance was measured in S1 and S2 under magnetic field applied from [001] to [100] and the magnetoresistance of the sample S1 was also measured between [001] and [110] in angular steps of 0.9 degrees.  The sample S3 has been measured in the angular range from [100] to [110]. In all cases the electrical current is applied along the [010] direction. 

\section{Results}

\subsection{Sample Characterisation}

\begin{figure}[!ht]
\begin{center}
\includegraphics[width=1\linewidth]{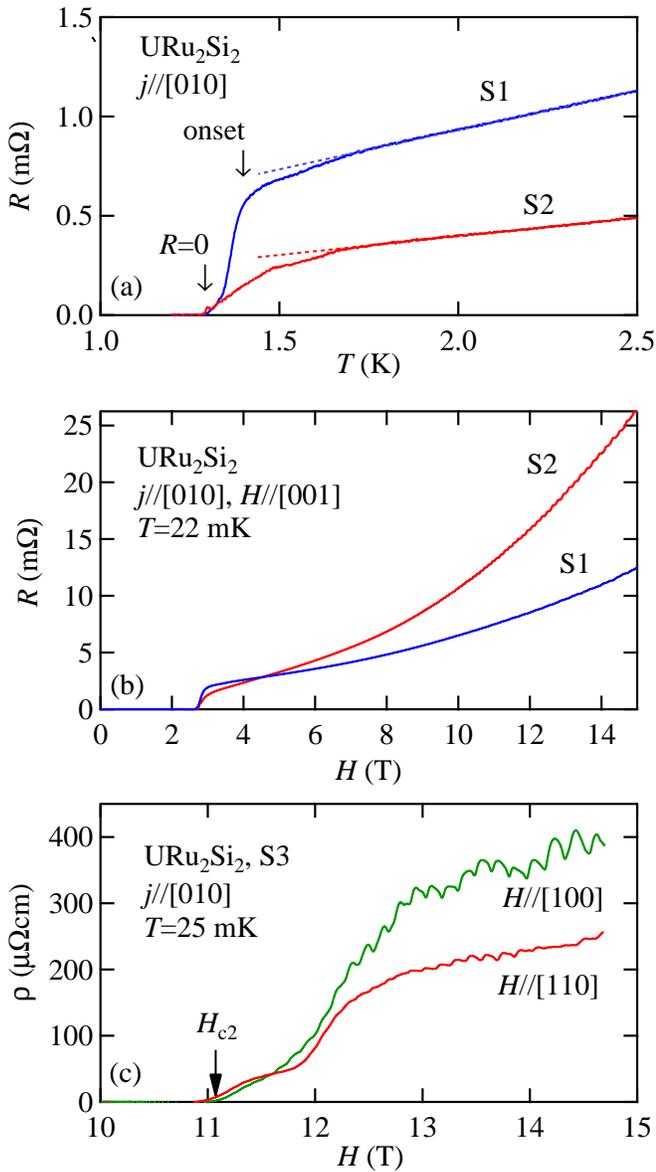}
\caption{(a) Temperature dependence of the electrical resistivity of samples S1 and S2. (b) Magnetoresistance at 22~mK for field applied along [001] of S1 and S2. (c) Magnetoresistance of S3 for field along [100] and [110] at 25 mK. The arrow indicates $H_{c2}$ for field along [100].}
\label{resistivityS1S2}
\end{center}
\end{figure}

The temperature dependence of the resistance of the two crystals studied in Grenoble is shown in Fig.~\ref{resistivityS1S2} (a). Both samples show zero resistance below $T_c=1.3$~K. The superconducting onset of S1 is at $T \approx 1.4$~K, while  sample S2 shows another pronounced kink at $T=1.5$~K. As indicated, in both samples, a tiny kink in $\rho (T)$ appears at $T = 1.7$~K indicating incipient superconducting fluctuations. In an extended temperature range from 1.7~K to 4~K the resistance can be parameterized with a power law and we find exponents $n=1.2$ for S1 and  $n=1.7$ for S2. Such a large variability of the temperature dependence of the resistivity above the superconducting transition has been reported previously \cite{Matsuda2011}, and it indicates the very strong sample dependence of the inelastic scattering in URu$_2$Si$_2$. The width of the superconducting transition of S1 defined as $\Delta T_{sc} = T_{sc}^{onset} - T_{sc}^{R=0} \approx 0.1$~K is comparable with previously studied high quality single crystals.\cite{Matsuda2008,Matsuda2011}

The magnetoresistance of the samples S1 and S2 is shown in Fig.~\ref{resistivityS1S2} (b) for field applied along the $c$ axis. 
 The flux-grown sample S2 shows a stronger magnetoresistance and the amplitude of the Shubnikov de Haas (SdH)  oscillations is larger for this sample indicating a higher average mean free path. Sample S1 was chosen for the study of the upper critical field $H_{c2}$ at different angles due to the sharper superconducting transition. S2 shows already strong superconducting fluctuations above the transition, and  the superconducting transition itself is also broader. In Oarai we measured the angular dependence of the SdH oscillations in S3 in the basal plane, fine turning the field from [100] to [110] in the field range from 10~T to 14.7~T. Figure \ref{resistivityS1S2} (c) shows the magnetoresistance of this sample for magnetic fields applied along [100] and [110].  Special attention has been taken to avoid a misorientation towards the $c$-axis, which nevertheless cannot be fully excluded. This sample shows extremely large quantum oscillations for $H \parallel [100]$.  The observed oscillations amplitude decreases when the field is applied along [110] mainly due to the fact that the current and field directions are 45 degree to each other and the magnetoresistance is between the transverse and longitudinal configuration for $H \parallel [110]$. In both directions we observe a distinct non-zero resistance between 11~T to 12~T. Zero resistance is observed below 11.05~T for $H\parallel [100]$ and 10.95 for $H\parallel [110]$, indicating that $H_{c2}$ is almost isotropic in the basal plane. \deleted{This small difference may be due to a tiny misalignment with respect to the $c$ axis.}  These values of the upper critical field $H_{c2}$ are lower than those previously reported.\cite{Brison1995, Ohkuni1999, Okazaki2008} Due to the strong quantum oscillations it is impossible to determine the width of the superconducting transition from field sweeps. 

\begin{figure}[!htb]
\begin{center}
\includegraphics[width=1\linewidth]{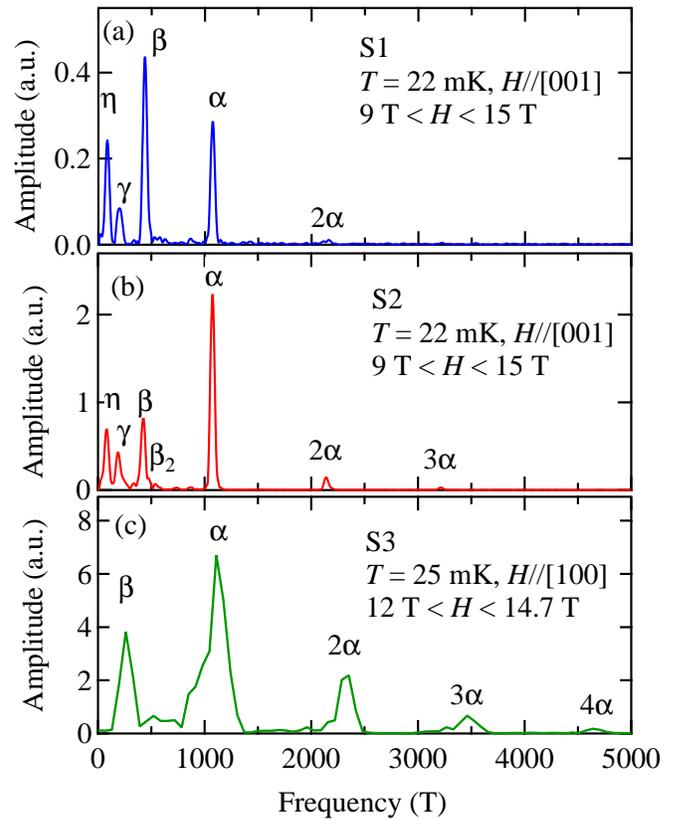}
\caption{(a-b) FFT spectra of the SdH oscillations at 22~mK in the field range from 9~T to 15~T along the [001] direction for S1 and S2. The quantum oscillation frequencies are indicated. $\beta_2$ corresponds to the spin split Fermi surface of the $\beta$ branch which appears due to the non-linear Zeeman splitting (see text). (c) FFT spectra for S3 at 25mK in the field range from 12~T to 14.7~T  applied along [100]. Up to four harmonics of the alpha branch are observed. The splitting of the $\alpha$ branch could not be resolved in this small field interval. } 
\label{FFT-spectra}
\end{center}
\end{figure}

Figure \ref{FFT-spectra} (a-b) shows the Fast Fourier Transformations (FFT) of the SdH oscillations at 22~mK for a field interval from 9~T to 15~T applied along the $c$ axis for samples S1 and S2. The spectra show four fundamental quantum oscillation frequencies in agreement with previous reports.\cite{Ohkuni1999, Shishido2009, Hassinger2010, Aoki2012} For the $\alpha$ branch of sample S2, one could detect up to the third harmonic in this field range. The relative amplitude of the different FFT frequencies changes between samples S1 and S2. While the $\beta$  frequency has the highest amplitude for S1,  the $\alpha$ frequency dominates the spectrum of S2.  Figure \ref{FFT-spectra} (c) presents the FFT spectrum of oscillations observed in S3 for field applied along [100]  in the field range from 12~T to 14.7~T. Up to four harmonics of the $\alpha$ branch are observed in this restricted field interval. The previously reported splitting of the $\alpha$ branch in the basal plane\cite{Ohkuni1999, Aoki2012}  could not be resolved in this small field interval but the asymmetry of the  FFT peak for the $\alpha$ frequency is an indication that the peak is a sum of different frequencies.

\subsection{Upper Critical Field}

\begin{figure}[!tb]
\begin{center}
\includegraphics[width=1\linewidth]{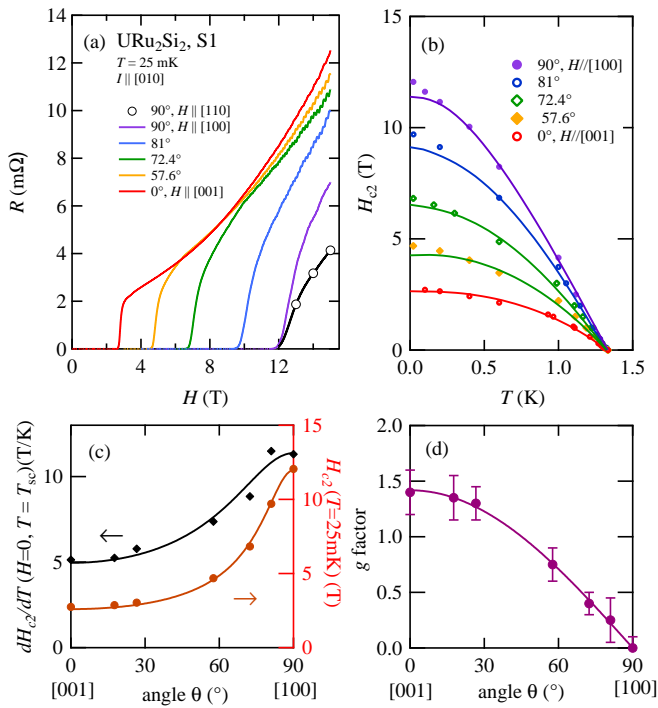}
\caption{(a) Magnetoresistance at $T=25$~mK for sample S1 for different field directions. (b) Temperature dependence of the upper critical field $H_{c2}$ determined by the criterion $R=0$. The solid lines are fits from the WHH model.\cite{Werthamer1966} (c) Angular dependence of the initial slope $-dH_{c2}/dT(T=T_{sc})$ and the low temperature value $H_{c2}(T=25~$mK) of the $H_{c2}$. Solid lines are guides to the eye. (d) Angular dependence of the $g$-factor deduced from the fits of the temperature dependence of $H_{c2}$. The solid line is a fit assuming an uniaxial $g$-factor anisotropy $g(\theta)=g_c \cos(\theta)$, with $g_c=1.4$.}
\label{Hc2}
\end{center}
\end{figure}

Figure \ref{Hc2}(a) displays the magnetoresistance at 25~mK for different angles measured on S1. To determine the upper critical field $H_{c2}$ the criterion $R=0$ has been chosen. The width of the transition is slighly increasing when the field is turned towards the basal plane. Close to [100], the onset of the superconducting state is no more clearly defined due to the oscillations of the magnetoresistance.  Similar to S3, the anisotropy  of $H_{c2}$ in the basal plane is very small. We find $H_{c2} = 12.05$~T for $H\parallel [100]$ and 11.95 for $H\parallel [110]$ at 25~mK. Figure~\ref{Hc2}(b) shows $H_{c2} (T)$ as a function of temperature for different magnetic field directions between [001] and [100]. The temperature dependence of $H_{c2}$ for [001] and [100] is in good agreement with previous studies.\cite{Brison1995,Ohkuni1999} A thermal conductivity study in URu$_2$Si$_2$ showed that the bulk upper critical field would be slightly higher than the resistive one.\cite{Okazaki2008} This small difference between the resistive and the bulk upper critical field will be neglected in the discussion. The angular dependence of the initial slope $-dH_{c2}/dT$ at $T=T_{sc}$, as well as that of $H_{c2}$ at 25~mK are represented in Fig.~\ref{Hc2}(c). Both are very anisotropic. \deleted{and we find} \added{The inital slope varies from  5.1~T/K to 11.3~T/K and the upper critical field from} $H_{c2}=2.75$~T to $12.05$~T at 25~mK, for field along [001] and [100] respectively. The initial slope of $H_{c2}$ at $T_{sc}$ allows an estimate  of the averaged anisotropy of the Fermi velocity ($H_{c2}' \propto \frac{T_{sc}}{v_F^2}$) which is given by $(v_{F}^{[001]}/v_{F}^{[100]})^2=(\frac{dH^{[001]}_{c2}}{dT}/\frac{dH^{[100]}_{c2}}{dT})^{-1}=1.5$. Here, $v_{F}^{[001]}$ and $v_{F}^{[100]}$ are the average Fermi velocity \added{of the quasiparticles} in the plane perpendicular to the direction of the magnetic field along $[001]$ and $[100]$, respectively. 

The temperature dependence of $H_{c2}$ has been calculated numerically based on the Werthamer, Helfand and Hohenberg (WHH) model within the weak coupling and clean limit \cite{Werthamer1966} with even parity pairing. \footnote{For simplicity the calculations are performed for a $s$ wave state. The exact form of the pairing symmetry has only minor corrections to the $T$ dependence.} Both, the paramagnetic and orbital limits are taken into account and the resulting fits are shown in Fig.\ref{Hc2}(b). The orbital limitation is controlled by the average Fermi velocity perpendicular to the applied magnetic field and it determines the initial slope $dH_{c2}/dT$ at $T_{sc}$,  while the paramagnetic limiting field is controlled by the electronic $g$-factor.\cite{Clogston1962} The WHH calculation reproduces the temperature dependence of $H_{c2}$ reasonably well, except at lowest temperatures, where the values from the experiment are slightly higher than the calculation.  

 The angular dependence of the $g$-factor extracted from these calculations of $H_{c2}$ is represented in Fig.~\ref{Hc2}(d). Under magnetic field along [100], the fit is  best for a complete absence of a paramagnetic limitation ($g^{[100]}=0$). Along [001] the $g$-factor  obtained by the fit is $g^{[001]}=1.4$. These results are in relatively good agreement with a previous similar study which yielded $g^{[100]}=0.2$ and $g^{[001]}$=1.9. \cite{Brison1995}   The angular dependence of the $g$-factor in Fig.~\ref{Hc2}(c) can be well fitted with $g(\theta)=g^{[001]} \cos(\theta)$, which corresponds to an Ising behavior of the quasiparticles. It is also consistent with the expected angular dependence of the paramagnetic limitation, when $g^{[100]}$ =0 (see Appendix of Ref.~\onlinecite{Brison1995}). Thus, from the upper critical field measurement, we can conclude that both the initial slope (and thus the Fermi velocity  of the quasiparticles) and the average $g$-factor of the dominant band for superconductivity are  anisotropic. \deleted{but with a much larger anisotropy of the $g$-factor.}
 \footnote{Another interpretation of the anisotropy of the upper critical field in URu$_2$Si$_2$ is based on the field dependence of the pairing interaction.\cite{Kusunose2012} However it needs a very low value of the coupling constant $\lambda=0.05$. This value would imply a difference of several order of magnitude between the characteristic temperature of fluctuations responsible for superconductivity and the superconducting temperature which seems unrealistic.}
\added{Essentially the $g$-factor in the basal plane determined from the superconducting upper critical field is close to zero and it is strongly increasing for fields close to the $c$ axis. The initial slope at $T_c$ (and thus the effective mass) is larger for field in the basal plane than for field along the $c$ axis. Importantly, the anisotropy of the effective mass from the initial slope is opposite to that determined  from the quantum oscillation, where the cyclotron masses for magnetic field applied along the $c$ axis is, for all orbits, larger than for field applied in the basal plane.\cite{Ohkuni1999,Hassinger2010,Aoki2012}. Thus, the anisotropy of the initial slope cannot be explained by the effective mass model with a single Fermi surface sheet.\cite{Morris1972}} This point will be discussed in section \ref{discussion_superconductivity}.

\subsection{Quantum Oscillations} 

\begin{figure}[!tb]
\begin{center}
\includegraphics[width=0.8\linewidth]{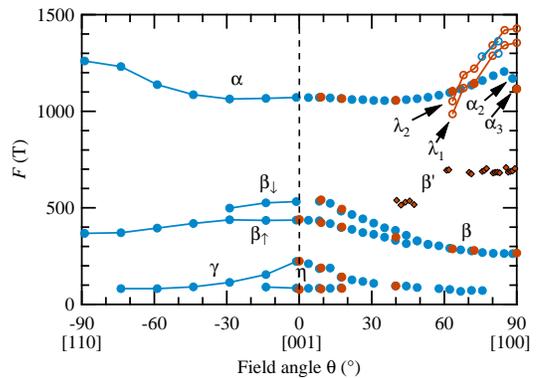}
\caption{Angular dependence of the SdH frequencies of URu$_2$Si$_2$. Blue and red symbols stand for samples S1 and S2, respectively. The $\eta, \gamma, \beta$, and the $\alpha$ branch are determined from the Fast Fourier transfortmations (FFT) on the interval 9--15~T. The heavy branch $\beta $',  which originates from the four-fold pockets of the $\beta$ Fermi surface, has been determined from an FFT in the field range 12--15~T. The light branches $\lambda_1$ and $\lambda_2$ have been determined at 600~mK in the field range from 9--15~T.
 }
\label{angular_dependence}
\end{center}
\end{figure}

The angular dependence of the quantum oscillation frequencies determined at 22~mK is plotted in Fig.~\ref{angular_dependence}. 
All previously reported branches have been observed,\cite{Bergemann1997,Ohkuni1999, Shishido2009, Hassinger2010, Aoki2012, Scheerer2014} except the light pocket  $\varepsilon$ which has been only reported in Ref.~\onlinecite{Shishido2009} to appear above 17~T. The nearly spherical Fermi surface pocket $\alpha$ and the $\beta$ Fermi surface are in good agreement with previous studies.  Close to [100] the $\alpha$ branch splits into \added{at least} three different branches in the basal plane.\cite{Ohkuni1999, Aoki2012} \added{The origin of the splitting is not fully understood, one proposal is that} it is due to a magnetic breakdown of a very tiny hour-glass Fermi surface at the $Z$ point of the Brillouin zone.\cite{Tonegawa2013} \added{As shown in Ref.~\onlinecite{Aoki2012}, the splitting is very sensitive to the perfect orientation in the basal plane. Under a small angle of 3 degree from the basal plane it is fully suppressed.} As already mentioned, this splitting is not resolved in this experiment \added{as the highest field in our experiment here is only 15~T}, but it is compatible with the broad asymetric FFT of sample S3 [see Fig.~\ref{FFT-spectra}(c)]. \added{Thus the analysis of the oscillation in the basal plane may not allow for a definive conclusion.}  

The $\beta$ Fermi surface consists of four pockets. As function of angle from [001] to [100] it splits into two branches: the $\beta$-branch and the heavy branch $\beta '$.\cite{Hassinger2010} The appearance of two frequencies $\beta$ and $\beta '$ for $H \parallel [100]$  proves that the pockets are located between the $\Gamma$ and $X$ points of the simple tetragonal Brillouin zone. Furthermore, the $\beta$ Fermi surface depends strongly on magnetic field.\cite{Aoki2012} \added{In agreement with the previous report,} we can resolve clearly a splitting  for the $\beta$ branch in two frequencies $\beta_\uparrow$ and $\beta_\downarrow$ in the angular range from [001] to 40$^\circ$ toward [110],  and from [001] to 15$^\circ$ toward [100], for $H>8$~T. (The assignment of the spin up and down branch will be justified below.) In this angle interval, the amplitude of the lower frequency $\beta_\uparrow$ is much stronger than that of $\beta_\downarrow$ and the amplitude of the FFT spectrum is only weakly modulated with  angle. However interferences between the signals from $\beta_\uparrow$ and $\beta_\downarrow$ can be observed on approaching [110] or [100]. It proves in agreement with the field dependence that the splitting of the $\beta$ branch near [001] is a spin splitting.  The strong field dependence confirms a non-linear Zeeman splitting. 

The angular dependence of the $\gamma$ and $\eta$ branches is similar to the previous report.\cite{Hassinger2010} We want to stress that the cross-section of the $\gamma$  orbit appears larger for field along the $c$ axis \added{and decreases in size to the basal plane}.  In difference, \added{all} band-structure calculations,\cite{Elgazzar2009, Oppeneer2010, Ikeda2012} \deleted{which} suggest an elliptical Fermi surface elongated along the $c$ axis.

 In addition, we have been able to determine the angular dependence of two light branches $\lambda_1$ and $\lambda_2$ at temperatures above 600~mK (see Fig.~\ref{angular_dependence}), when the amplitude of the heavy branches is strongly suppressed. These branches have been observed in previous experiments in pulsed magnetic fields.\cite{Scheerer2014} From the temperature dependence of the amplitude, which has been measured up to 1 K, we determine the effective masses of these light branches to $m^\star_{\lambda_1} = 1.4 m_0$ and $m^\star_{\lambda_2} = 2.1 m_0$. Band structure calculations do not predict such light frequencies. They may correspond to the light bands $F$ and $G$ observed in cyclotron resonance experiments.\cite{Tonegawa2012, Tonegawa2013} 

The spin degeneracy of  the conducting electrons is lifted in an applied magnetic field leading to an energy difference between the spin up and spin down electrons which is given by the Zeeman term $\Delta E = \frac{1}{2}g \frac{e\hbar}{m^\star} H$. \added{The Fermi surface splits in spin-up and spin-down sheets.} 
\deleted{The spin-slitting of the Fermi surface  in spin up and down electrons can give rise to interferences, leading to a modulation of the amplitude of the quantum oscillations.  }
\deleted{The phases of quantum oscillations from spin up and spin down Fermi-surface pockets in a paramagnet with a linear Zeeman splitting are given by }
\begin{equation*}
 \label{phaselinearup}
\deleted{\phi_\uparrow  =\phi_0 + \pi\frac{m_\uparrow^\star g}{2 m_0}}
\end{equation*}
\begin{equation*}
\deleted{
\phi_\downarrow  = \phi_0 - \pi\frac{m_\downarrow^\star g}{2 m_0}\\}
\label{phaselineardown}
\end{equation*}
\deleted{with the spin dependent effective masses $m^\star_{\uparrow}$ ($m^\star_{\downarrow}$) for spin up (down). }
\added{The effect of this spin splitting is equivalent to a phase difference of $\phi = 2 \Delta E = \pi g \frac{m^\star}{m_0}$ between the oscillations coming from the spin up and spin down electrons and can give rise to interferences, leading to modulations of the amplitude of the quantum oscillations. This simple approach for free electrons neglects all field dependences of the cyclotron orbits, the effective mass and also the effective spin splitting $g$ factor.}

The quantum oscillation frequencies $F$ are related to the extremal cross-section ${\cal A}$ of the Fermi surface by the Onsager relation $F = (\hbar/2\pi e){\cal A}$. However, the frequency $F_{obs}$, which is measured in the experiment at a finite field, is related to the true quantum oscillation frequency by $F_{obs} (H) = F_{true} - \frac{dF_{true}}{dH}$.\cite{Ruitenbeek1982} What is measured is the so-called back-projected frequency to zero field. Thus, if the observed \deleted{quantum oscillation} frequency is field independent \added{the true frequency increases linearly with field and thus} the Zeeman-splitting of the Fermi surface is also linear in field. 
In the case that the back-projection to zero field of the frequencies of the spin-up and spin-down quantum oscillations, \deleted{are equal  in absence of any spontaneous magnetization} and the effective masses and mean free path of the quasiparticles \deleted{with different spin direction are equal}\added{do not depend on the spin direction}, the angular dependence of the amplitude of the first harmonics of the quantum oscillations can be described by Eq.~\ref{spinzeroang}.
 The amplitude of the quantum oscillations vanishes when the spin-splitting damping factor $\cos(\pi g m^\star/2m_0)$ is zero, i.e.~when $g(m^\star/m_0) =2n -1$ is an odd integer.
 
 However, \added{if} the observed frequency $F_{obs}$ \deleted{can be}\added{is} field dependent, $F_{true}$ has a non-linear field response. In this case the observed frequencies $F_\uparrow$ and $F_\downarrow$ of spin-up and spin-down Fermi surfaces are not identical and the damping factor does not vanish. \added{Due to the non-linear response, the back-projected frequencies for spin-up and spin-down are not identical and two frequencies are observed. }   \added{Generally, in heavy fermion systems the effective mass of the quasiparticles is expected to be spin dependent \cite{Spalek2006, Kaczmarczyk2009} and such a spin dependence has been experimentally observed.\cite{Sheikin2003, McCollam2005} } \footnote{\added{The detection of spin-splitting zeros excludes the presence of any spontaneous magnetization, as in that case the orbits of spin-up and spin-down electrons have different sizes.}}
\added{In addition, also the effective $g$ factor can be field dependent. However, the experimental observation of a field dependent $g$ factor is rare.\cite{Harrison2015}
}

\begin{figure}[!hb]
\begin{center}
\includegraphics[width=1\linewidth]{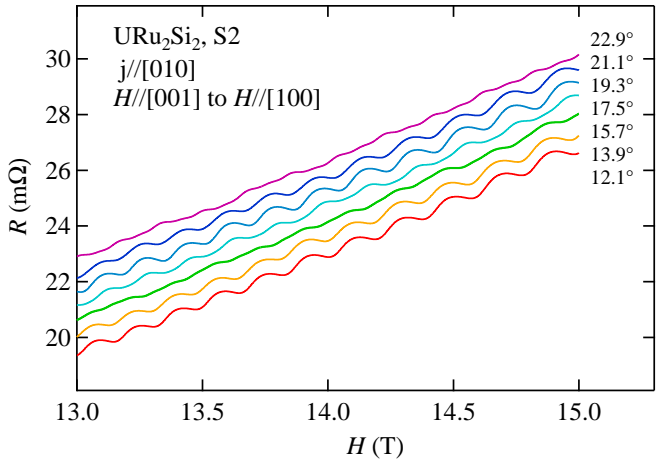}
\caption{Resistance measured on S2 for different angles from [001] to [100]. The curves are vertically shifted of 0.5 m$\Omega$ for clarity. The amplitude of the SdH oscillations change with increasing angle and is minimal for $\phi = 15.7^\circ$.  A phase shift of 180$^\circ$ occurs indicating a spin-splitting zero. 
}
\label{rho_osc}
\end{center}
\end{figure}

\begin{figure}[!h]
\begin{center}
\includegraphics[width=1\linewidth]{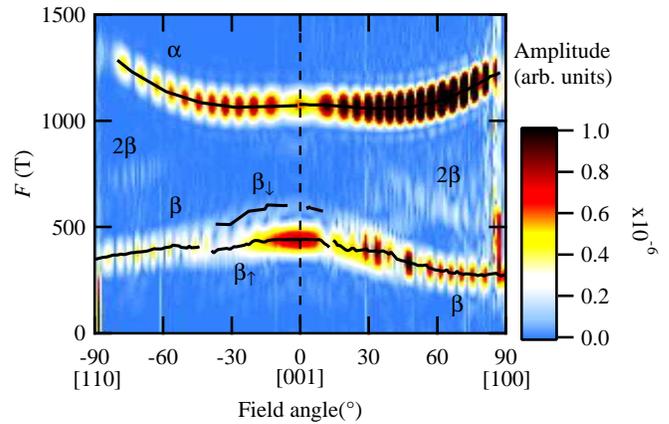}
\caption{FFT spectra of quantum oscillations at $T=22~$mK in the field range from 12~T to 15~T  as a function of angle for S1. The color code corresponds to the amplitude of the FFT spectra. The solid lines show the angular dependence of $\alpha$ and $\beta$ frequencies. In the vicinity of [001] the signal of the $\beta$ branch splits into $\beta_\uparrow$ and $\beta_\downarrow$. }
\label{color}
\end{center}
\end{figure}

The magnetoresistance at 22 mK measured in S2 is represented for different field angles from 12.1$^\circ$ to 22.9$^\circ$ from [001] to [100] in Fig.~\ref{rho_osc}. The SdH oscillations from the $\alpha$ branch are clearly resolved. The quantum oscillation amplitude decreases from 12$^\circ$ to nearly $16^\circ$ and increases for larger angles. A phase shift of 180$^\circ$ can be observed between oscillations observed for angles slightly below and above 16$^\circ$. This is a clear indication for the appearance of a spin-splitting zero. 

Figure~\ref{color} shows a contour plot of the amplitude of the FFT spectra calculated in the field interval 12$~$T--15$~$T of the quantum oscillations at $T=22~$mK as a function of angle for sample S1. The horizontal and vertical axes correspond to the field angle and the oscillation frequency respectively. The solid lines in Fig.~\ref{color} gives the angular dependence of the SdH frequencies in this field range. In this color plot the appearance of spin zero is clearly observed for the $\alpha$ and $\beta$ branches. Next we will discuss the oscillation of the amplitude for the different branches in detail.

\begin{figure}[!h]
\begin{center}
\includegraphics[width=1\linewidth]{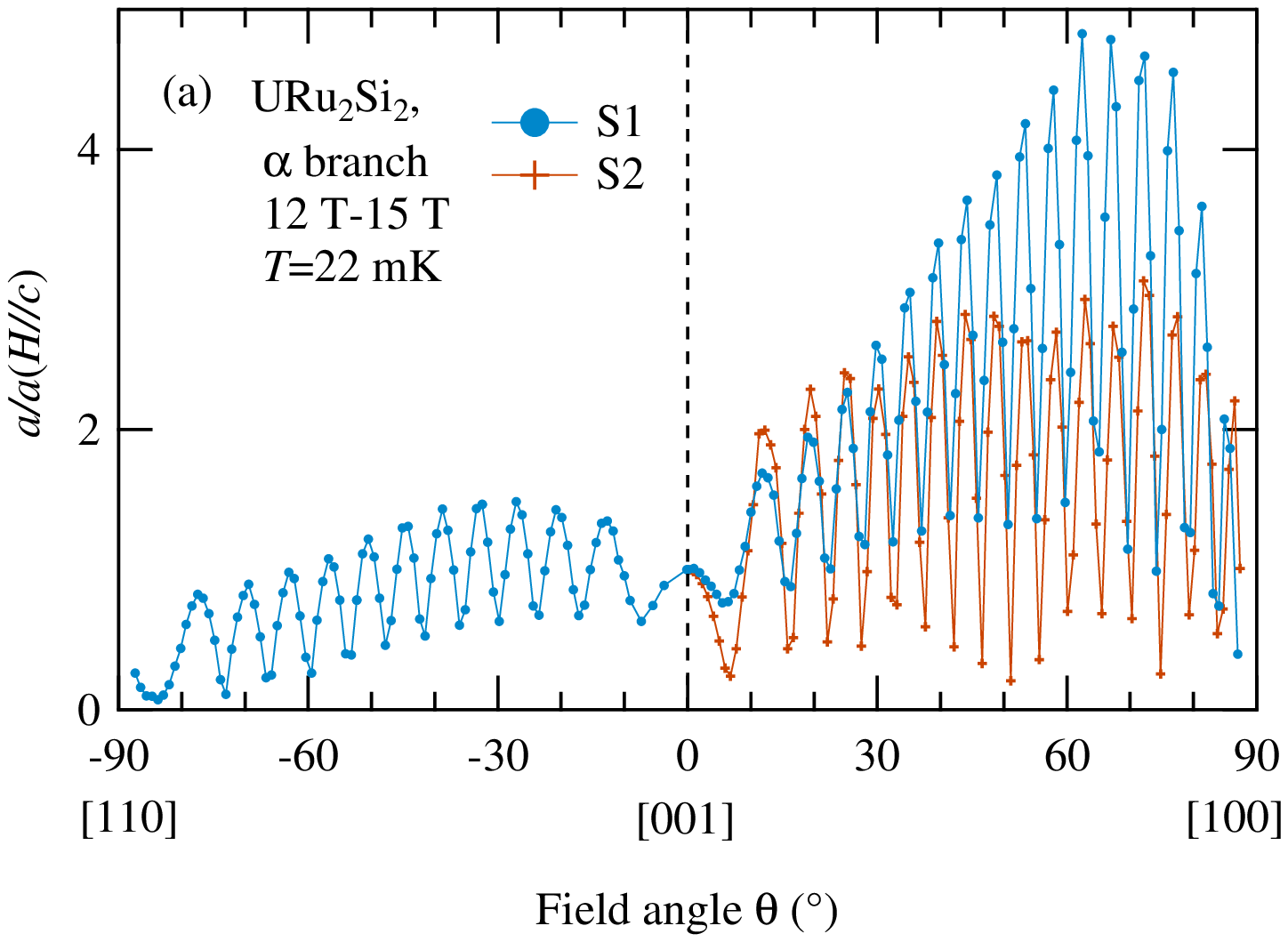}
\includegraphics[width=0.9\linewidth]{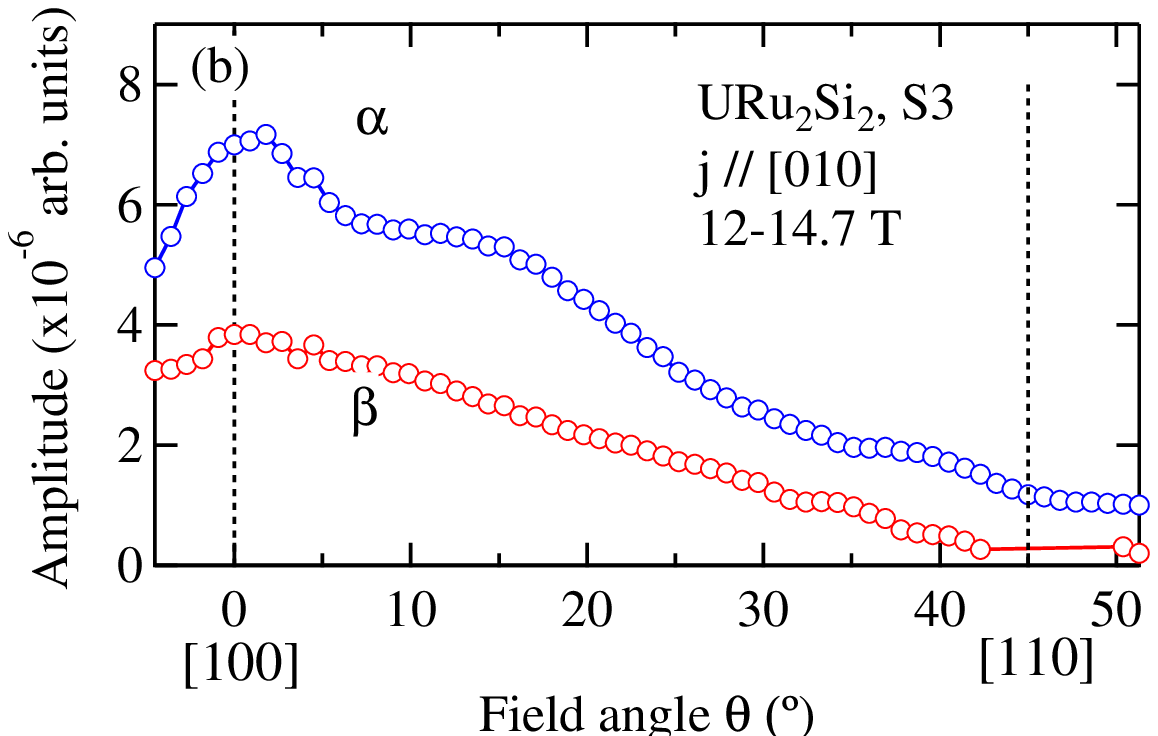}
\caption{(a) Renormalized amplitude of quantum oscillations from the $\alpha$ orbit at 22$~$mK on the field interval 12$~$T-15$~$T as a function of the field direction for S1 and S2. (b) Angular dependence of the amplitudes of the $\alpha$ and $\beta$ orbits in the basal plane observed in S3.
} 
\label{interferences}
\end{center}
\end{figure} 

Figure \ref{interferences}(a) displays the angular dependence of the oscillations amplitude for the $\alpha$ Fermi-surface pocket from [001] to [110] and from [001] to [100]  in the field interval 12~T--15~T measured on S1 (blue circles) and S2 (red crosses). The amplitude is normalized to the value at $H \parallel [001]$. The amplitude oscillates very strongly with the field angle. In the field interval 6~T -- 9~T similar oscillations of the amplitude have been observed which indicates that they are not field dependent.  Comparable oscillations of the de Haas van Alphen (dHvA) amplitude from the $\alpha$ pocket have already been reported in Ref.~\onlinecite{Ohkuni1999}. While Ohkuni et al.~observed 16 spin-splitting zeros between [001] and [100], both samples in our measurements show 17 zeros. This difference \deleted{may}\added{can} be \deleted{due to}\added{explained by} a slight misalignment in the previous experiment\cite{Ohkuni1999} around an axis transverse to the rotation axis. In difference, when turning the field from [001] to [110] we observe only 12 spin-splitting zeros. Note that the amplitude does not vanish completely at the spin-splitting zeros. Already the previous data of Ohkuni et al.~\cite{Ohkuni1999} showed a similar behaviour of washed out spin-splitting zeros.  This can be explained by small differences in the frequencies or in the effective masses of spin-up and spin-down bands, which are to small to be resolved in our experiment. \added{Generally, a strong spin dependence of the effective mass is expected in heavy fermion systems.\cite{Korbel1995, Spalek2006}} In other systems, where spin-splitting zeros have already been reported, such finite values of the amplitude had been reported.  In Sr$_2$RuO$_4$ it has been argued that the washed-out spin-zeros are due to a different warping for the spin-up and down- parts of the cylindrical Fermi surfaces.\cite{Bergemann2003}  The variations between different samples may be due to a different amount of impurities. Note that we already observed differences in the relative size of the FFT amplitudes of S1 and S2, which also indicates differences in the Dingle temperature of the various orbits. 

\added{As discussed above,} in the basal plane the $\alpha$ branch splits in \deleted{at least three} different frequencies \cite{Ohkuni1999, Aoki2012}. \added{Close to [110] three frequencies have been observed with effective masses of 9.7 $m_0$,  12 $m_0$, and 17 $m_0$ which change little as function of angle in the basal plane. In a limited angular range Ohkuni et.~al reported that branch $\alpha$ is even four-fold split. } \deleted{A small misalignment of three degrees is be enough to suppress the splitting.}\cite{Aoki2012} \added{Here, we do not see any splitting of the frequencies in the basal plane, contrary to the angular dependence of the cyclotron resonance frequencies reported in Ref.~\onlinecite{Tonegawa2012, Tonegawa2012}.} In Fig.~\ref{interferences} (b) we show the angular dependence of the oscillation frequency observed in the field range from 12~T to 14.7~T. \added{As already shown in Fig.~\ref{FFT-spectra} (c)} this splitting is not resolved in the present experiment due to the small field interval \added{from 12~T to 14.7~T}.  Thus it is not surpriging that no spin-splitting zero is observed in the basal plane when turning the magnetic field from [100] to [110]. The decrease of the amplitude from [100] to [110] for both orbits is due to the change of the current direction with respect to the magnetic field from  a transverse configuration (current perpendicular to the field) to 45 deg.~with respect to the field axis. \added{In any case, also in our previous experiment \cite{Aoki2012} we did not see any indications for any spin-splitting zero.} 

\begin{figure}[!h]
\begin{center}
\includegraphics[width=1\linewidth]{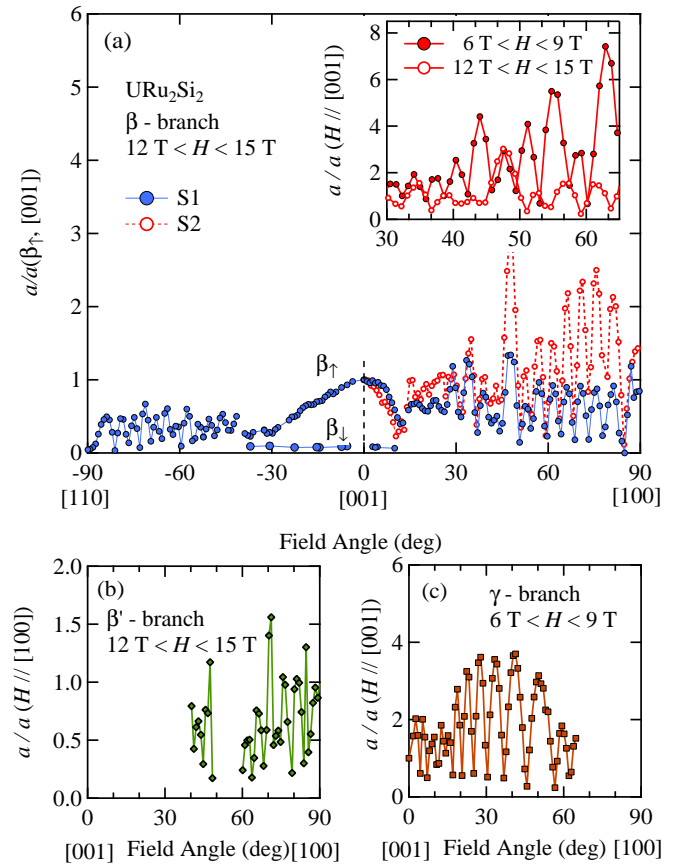}
\caption{(a) Normalized amplitude of quantum oscillations from the $\beta$ branch as function of field angle. In the field range close to [001] spin-zero splitting of the $\beta$ branch appears and the interference of spin up and spin down electrons is no more perfect. In the angular range between [001]  to 40$^\circ$ to [110] and to 15$^\circ$ to [100] we clearly observe two frequencies, $\beta_\uparrow$ and $\beta_\downarrow$. \added{The inset shows the angular dependence from 30 deg to 65 deg of amplitude of the $\beta _\uparrow$ oscillation in different field ranges for S2}. (b) and (c) shows the normalized amplitudes of the $\beta '$ and $\gamma$ branches measured in S1, respectively. For the $\gamma$ branch the field interval has been 6~T -- 9~T.
} 
\label{interferences_beta}
\end{center}
\end{figure} 

The angular dependence of the oscillation amplitude from the $\beta$ branch is represented in Fig.~\ref{interferences_beta}(a). It is determined from the FFT spectra in the field range from 12~T to 15~T. Near to [001], only very weak oscillations of the amplitude have been observed. This is due to the  spin-splitting of the $\beta$ frequency under magnetic field \added{(see Fig.~\ref{angular_dependence} and also Fig.~6 of Ref.~\onlinecite{Aoki2012}) }. \added{The field dependence of the observed quantum oscillations of branch $\beta$ can be interpreted as non-linear field dependence of the minority spin-down Fermi surface which shrinks with increasing magnetic field and gives rise to a strong increase of the effective mass, as $m^\star  = \frac{\hbar^2}{2\pi}\frac{\partial {\cal A}(k_H)}{\partial E}|_{E_F}$, where ${\cal A}(k_H)$ is the cross-sectional area of the Fermi surface which is perpendicular to the field and $k_H$ is the wave number along the field direction.\cite{Onuki1995} The effective mass of the spin-minority band increases up to 40$m_0$.   This non-linear field dependence of the quantum oscillation frequencies is the consequence of the polarization of the small and heavy electron-like $\beta$ Fermi-surface pocket under magnetic field along the easy magnetization axis.\cite{Altarawneh2011, Aoki2012, Scheerer2014}
Thermopower measurements in URu$_2$Si$_2$ under magnetic field along the $c$-axis show a minimum at $H_m=11~$T at low temperature \cite{Malone2011, Pourret2013}, which also indicates an evolution of the Fermi surface with the magnetic field. Further field-induced Fermi surface changes inside the hidden order state have been detected at higher magnetic field by Hall effect,\cite{Shishido2009} thermoelectric power,\cite{Malone2011, Pourret2013} and quantum oscillations.\cite{Altarawneh2011, Aoki2012, Scheerer2014} 
However, \deleted{we want to mention that} all these Fermi surface changes inside the hidden order state has almost no feedback on the measured macroscopic magnetization which increases almost linearly with field up to $H_c \approx 35$~T, where the hidden order is suppressed, and the magnetization shows a first order metamagnetic jump.\cite{deBoer1986, Sugiyama1999, Scheerer2012} Only the NMR Knight shift shows a tiny increase at 23~T,\cite{Sakai2014} where a new quantum oscillation frequency appears.\cite{Shishido2009, Aoki2012}
}

In the angular range further away from [001], the spin-splitting is no more resolved (see Fig.~\ref{angular_dependence})and the frequencies of spin-up and spin-down Fermi surfaces coincide. While the amplitude of $\beta_\uparrow$ oscillations is maximum at [001], the amplitude of $\beta_\downarrow$ oscillations is much smaller and nearly constant with angle. The amplitude of the $\beta$ oscillations shows 11 spin-splitting zeros between [110] and 40$^\circ$ from [001] and 13 spin-splitting zeros between [100] and 15$^\circ$ from [001]. Between [001] and [100] both samples show the same number of spin-zero. However, spin-splitting zeros are more clearly resolved in sample S1. In this sample the amplitude of the $\beta$ frequency for $H\parallel [001]$ is larger than that of the $\alpha$ branch but, compared to S2, the oscillation amplitude is lower. The oscillation amplitude (in both samples) does not vanish completely at the spin-splitting zeros. Again, it must come from the incomplete cancellation of spin-up and spin-down oscillations due to their amplitude difference and their small frequency and effective mass difference. No spin-splitting zero is observed in the basal plane, when turning the magnetic field from [100] to [110], see Fig.~\ref{interferences}(b), but the amplitude decreases smoothly due to the change in the magnetoresistance. 
\added{The inset in Fig.~\ref{interferences_beta}  shows the angular dependence of the oscillations in the field range from 6~T -- 9~T from 30$^\circ$ -- 65$^\circ$ from [001] in the (010) plane. Below 9~T, no spin-splitting of the $\beta$ branch is observed. Remarkably,  for $6~T < H < 9~T$ the spin-splitting zeros are closer to each other with 8 spin zero between 30$^\circ$ and 65$^\circ$ against 7 for the field interval 12~T --15~T. The non-linear expansion of the spin majority Fermi surface leads to a non-linear Zeeman effect and to a reduction of the number of spin-splitting zeros under field.This is different than for branch $\alpha$ where the same number of spin-zeros had been observed, independent of the magnetic field range.  }

\deleted{The phase of $\beta_\uparrow$ oscillations is plotted in Fig.~\ref{interferences_beta}(c). It was determined by fitting the SdH oscillations with a sum of two sinusoids corresponding to oscillations from the $\beta_\uparrow$ and $\alpha$ orbits, which are the dominant frequencies. The phase increases continuously with angle from [001] to [100] or [110]. The phase should follow Eqs.~(\ref{phaselinearup},\ref{phaselineardown}) for spin-up or spin-down, respectively.} 

\deleted{
From band-structure calculations we know that the $\beta$ pocket is electron-like.\cite{Elgazzar2009, Ikeda2012} As the $g$-factor is expected to be maximum at [001], the lowest of the two frequencies from the $\beta$ branch corresponds to the minority spin band $\beta_\uparrow$. This allows us to interpret the field dependence of the quantum oscillations frequencies and effective masses published in Ref.~\onlinecite{Aoki2012} and we can conclude that the expansion of the spin majority Fermi-surface pocket for the $\beta$ Fermi surface saturates and this pocket gets  heavier under magnetic field along the easy magnetization $c$ axis. Such a non-linear expansion of the spin majority Fermi surface leads to a non-linear Zeeman effect and to a reduction of the number of spin-splitting zeros under field. 
This non-linear field dependence of the quantum oscillation frequencies is the consequence of the polarization of the small and heavy electron-like $\beta$ Fermi-surface pocket under magnetic field along the easy magnetization axis.\cite{Altarawneh2011, Aoki2012, Scheerer2014}
Thermopower measurements in URu$_2$Si$_2$ under magnetic field along the $c$-axis show a minimum at $H_m=11~$T at low temperature \cite{Malone2011, Pourret2013}, which also indicates an evolution of the Fermi surface with the magnetic field. Further field-induced Fermi surface changes inside the hidden order state have been detected at higher magnetic field by Hall effect,\cite{Shishido2009} thermoelectric power,\cite{Malone2011, Pourret2013} and quantum oscillations.\cite{Altarawneh2011, Aoki2012, Scheerer2014}
}

The amplitude of the $\beta'$ orbit quantum oscillations is represented as a function of angle between [001] and [100] in Fig.~\ref{interferences_beta}(b). It could be determined only in sample S2 in the interval 12 -- 15$~$T and $\beta'$ could not be resolved below 40$^\circ$ due to the proximity \added{of its oscillation frequency} with $\beta_\uparrow$  and also not between 50$^\circ$ and $60^\circ$ due to the proximity to the \added{frequency} of the second harmonic from $\beta$ orbit. It shows three spin-zero between 40$^\circ$ and 50$^\circ$ and seven from $60^\circ$ to $90^\circ$. 

The $\gamma$ Fermi-surface pocket is a small ellipsoid with $F_\gamma=200~$T along [001] and $F_\gamma=70~$T in plane.\cite{Aoki2012} Its frequency is too small to be resolved in the interval 12~T -- 15~T, so this pocket was studied only in the interval 6~T -- 9~T. The oscillation amplitude in S2 is represented as a function of the angle from [001] toward [100] in Fig.~\ref{interferences}(c). Twelve spin-splitting zeros are observed up to 65$^\circ$. For higher angles the signal of the $\gamma$ branch cannot be followed in this field range due to the superconducting transition.

\section{Discussion}

\subsection{Analysis of the $g$-factor}
 
According to Eq.~(\ref{spinzeroang}), the amplitude of the quantum oscillations vanishes if $m^\star g /m_0 = 2n -1$ with \deleted{$n = 1, 2, ...$} \added{with $n$ being the number of the spin-splitting zero}. \added{The argument of the cos-term of the spin-factor $m^\star g/2 m_0 $ is an integer number at each maximum of the amplitude in the angular dependence. Thus we can  determine the value of $m^\star g/2m_0$ only up to an integer number $k$. } Generally, we can expect the appearance of spin-splitting zeros with field angle, if the $g$-factor or the effective mass are highly anisotropic and $g$ or $(m^\star/m_0)$ are large enough. From the spin-splitting zeros, only the product $g (m^\star /m_0)$ can be determined \added{and the effective mass $m^\star$ has to be determined from the temperature dependence of the oscillations.}

\begin{figure}[!tb]
\begin{center}
\includegraphics[width=1\linewidth]{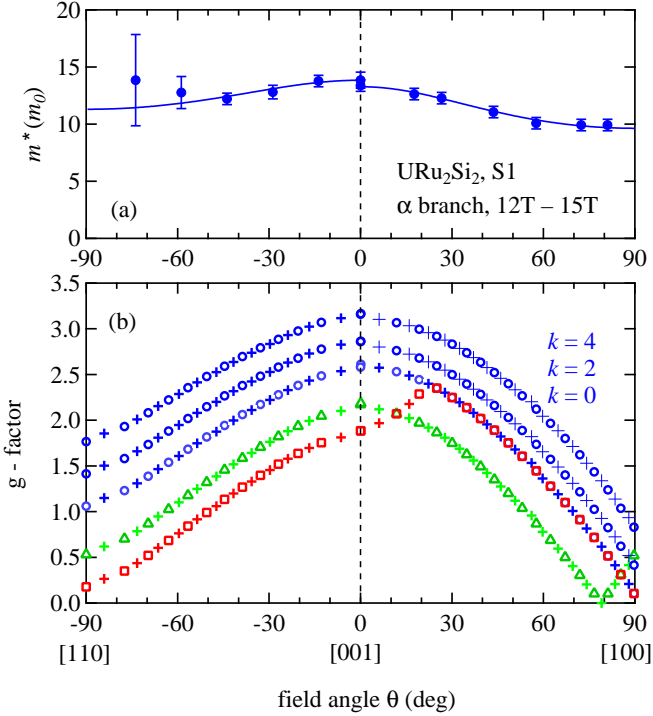}
\caption{(a) Angular dependence of the effective mass of the $\alpha$ branch in the field range from 12~T -- 15~T. (b) Angular dependence of the $g$-factor for the $\alpha$ branch of URu$_2$Si$_2$ for different values from $k = 0, 2,$ and 4. Crosses are from the spin-splitting zero, circles from the maxima of the amplitude of the SdH amplitude. The red and green curves give possible scenarios } 
\label{g-factor}
\end{center}
\end{figure}

For the $\alpha$ Fermi surface the effective mass $m^\star$ is rather isotropic. We have determined the effective mass $m^\star$ for different directions and find $m^\star_{[001]} = 13.3 m_0$,  $m^\star_{[100]} = 9.7 m_0$, and $m^\star_{[110]} = 11.3 m_0$ for fields applied along [001], [100], and [110], respectively. \added{As discussed above, in the basal plane the $\alpha$ branch is splitted in at least three branches.} The effective mass evolves smoothly between these principal axes [see Fig.~\ref{g-factor}(a)]. 
 \deleted{ As the $g$-factor is determined from the cosine term, } Different solutions exist for $g$ and the determination is not unique. Figure \ref{g-factor}(b) shows possibilities for the angular dependence of the $g$-factor for the $\alpha$ branch of URu$_2$Si$_2$ from the spin-damping factor depending on the choice of $k$ (blue symbols). We assume that the $g$-factor should be largest along [001] and the value $g (m^\star/m_0)$ changes monotonously as a function of field angle \added{and we choose $k$  as the value of $m^\star g/ 2m_0$ at the closest amplitude maximum from [100]. }

For $k=0$, the data suggest a strong anisotropy of $g$  from [001] to the basal plane varying from $g_\alpha^{[001]} \approx 2.5$ to $g_\alpha^{[100]} \approx 0$ along the $a$ direction. However, as we have only observed 12 spin-splitting zeros when turning the angle from [001] to [110] and the effective mass does not change significantly between [100] and [110], we find $g_\alpha^{[110]} \approx 1.1$ along [110], i.e.~it is not vanishing, \added{but would indicate a large anisotropy of $g$ in the basal plane. The main difference between the curves for the different values of $k$ is a vertical shift, so the variation of the effective mass with angle gives only a small correction.} 

\added{Nevertheless, as we observed 17 spin zero from [001] to [100], but only  12 from [001] to [110], 5 spin-splitting zero have to be observed in the basal plane.    }
\deleted{The difference between $g$ along [100] and [110] would imply the observation of at least five spin-splitting zeros in the basal plane. }
This is at odds with the variation of the SdH amplitude in the basal plane shown in Fig.~\ref{interferences}(b), and also with our previous high field experiment \cite{Aoki2012} \added{and that of Ohkuni et al. \cite{Ohkuni1999}}\deleted{where no spin-splitting zero has been observed.} 
\added{As discussed above, the splitting of branch $\alpha$ could not be observed in our present experiment with maximal field of 15~T.  Assuming the three orbits of the $\alpha$ branch ($\alpha_1, \alpha_2$ and $\alpha_3$)  are spin degenerated, the observed oscillation amplitude would originate from the interference between oscillations of the six orbits. It explains why it is nearly constant with the magnetic field angle between [100] and [110]. In the  previous experiment in the field range 12~T to 30~T,\cite{Aoki2012} where the splitting of the $\alpha$ branch has been resolved, only for the branch $\alpha_2$ a spin-splitting zero may occur between [100] and [110]. On the contrary as we have observed 17 in the (010) plane and 12 in the (110) plane, which means that the phase of the oscillations change by $17\pi$ and $12\pi$ respectively, suggests the occurrence of five spin zeros in plane, the phase of the oscillations change by $5\pi$, under the assumption that $g m^\star$ has a monotonous evolution from the $c$ axis to the basal plane. If we allow a non-monotonous variation of the $g$-factor, possible solutions could be a maximum (red curve in Fig.~\ref{g-factor}) or a minimum of the $g$-factor (green curve).\footnote{Of course other solutions may be possible too. } Only if we take into account a non-monotonous variation of $g$, a self-consistent solution for the $\alpha$ branch can be found from our data.\footnote{These is only valid under the assumption that there is no spin-splitting zero in the basal plane. Nevertheless, there is no other experiment that supports a non-monotonous variation of a physical property in the (010) plane that it is difficult to imagine that $g$ has a maximum near 30 deg from the $c$ axis. } }

\added{The cyclotron resonance experiment reported in Ref.~\onlinecite{Tonegawa2012} showed an unusual splitting of the sharpest observed resonance line which is assigned to the $\alpha$ Fermi surface sheet under in-plane magnetic field rotation from [100] to [110] in the basal plane.  The observed splitting is explained  by a domain formation  which breaks the tetragonal symmetry and accounts for by the in-plane mass anisotropy which has heavy (hot) spots only near the orbit for $H \parallel [110]$ and $H \parallel [\overline{1}10]$. This domain formation suggests to explain the observed breaking the tetragonal symmetry in the basal plane.\cite{Okazaki2011} However, the recent high resolution X-ray experiment \cite{Tabata2014, Amitsuka2017,Choi2018} and also NMR results\cite{Walstedt2016,Kambe2018}  do not confirm the previously reported tetragonal symmetry breaking.\cite{Okazaki2011, Tonegawa2012, Tonegawa2013, Tonegawa2014, Riggs2015} }
\deleted{In addition, the upper critical field $H_{c2}$ is isotropic in the basal plane. As indicated in Fig.~\ref{g-factor}(b) the observed difference between $g_\alpha^{[110]}$ and $g_\alpha^{[100]}$ is decreasing for higher values of $n$, e.g.~for $n=20$ the difference for $g$  along [110] and [100] would be only 0.44, but the $g$-factor would vary between 4 and 5.5 between the basal plane and the $c$ direction, respectively. Such large $g$-factors have been observed in other intermetallic systems like YRu$_2$Si$_2$ with $g=6$.\cite{Ikezawa1995} 
}
\deleted{If we ignore the multiband character, such large values for the $g$-factor are difficult to concile with the superconducting upper critical field $H_{c2}$. Taking account of the multiband character of URu$_2$Si$_2$, the ``contradiction'' between a value and the anisotropy of the paramagnetic limitation of $H_{c2}$, and that the anisotropy of the $g$-factor of the $\alpha$ branch implies that superconductivity is not governed by the pairing of electrons from the $\alpha$ branch. Furthermore, such a large, isotropic $g$-factor from the $\alpha$ branch also cannot account for the large anisotropy of the spin susceptibility in the normal state,\cite{Hattori2016}  as discussed in Ref.~\onlinecite{Altarawneh2012}. However, it should be mentioned that the $\alpha$ branch, even if it is the largest Fermi-surface pocket, contributes to only $\gamma_\alpha \approx 6.1$~mJ/(mol K$^2$) to the total specific heat coefficient of $\gamma \approx 55$~mJ/(mol K$^2$).}

\begin{figure}[!h]
\begin{center}
\includegraphics[width=1\linewidth]{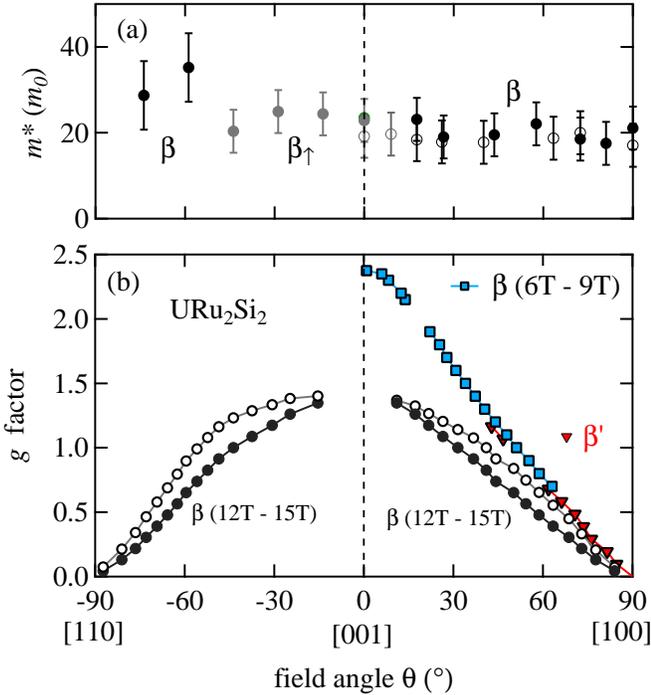}
\caption{(a) Angular dependence of the effective mass of the $\beta$ branch in the field range 12~T--15~T. Closed (open) circles are for S1 (S2), respectively. Grey symbols mark the effective mass for the spin-split spin-up branch $\beta_\uparrow$. (b) Angular dependence of the $g_\beta$ factor for the $\beta$ branch of URu$_2$Si$_2$ for the most anisotropic scenario with $k$=0. Other solutions can be found by shifting the curves upwards. Black circles give the $g$-factor in the field range 12~T -- 15~T with a constant mass as in determined in the field range from 12~T--15~T [see upper panel (a)]. Open circles give $g$ assuming the angular dependence of the effective mass in Ref.~\onlinecite{Aoki2012}. Blue squares show the result for  the $g$-factor in a lower field range from 6~T -- 9~T. Red triangles indicate the angular dependence of $g$ for the heavy orbit $\beta '$ from [001] to [100].
} 
\label{g-beta}
\end{center}
\end{figure}

\deleted{Next we will concentrate on the heavy $\beta$ pocket. This branch}  The heavy $\beta$ pocket shows a very strong field dependence above 8~T for $H\parallel [001]$.  The observed SdH frequency splits under magnetic field as a consequence of the non-linear Zeeman effect.\cite{Aoki2012, Scheerer2014} Therefore the $g$-factor was calculated in the field interval 6$~$T -- 9$~$T with reduced effect of the non-linear field splitting, and for comparison, in a higher field range from 12~T -- 15~T. In Fig.~\ref{g-beta}(a) we plot the angular dependence of the mass of the $\beta$ branch determined for samples S1 and S2. We observe an almost constant effective mass for the $\beta$ branch within the error bars. Therefore we use $m^\star \approx 21 m_0$, independent of angle. In difference, our previous data showed that the effective mass of the $\beta$ branch shows a rather strong angular dependence changing from $m^\star_\beta \approx 23 m_0$ for field along [001] to $m^\star_\beta \approx 13.5 m_0$.\cite{Hassinger2010, Aoki2012}  This is probably due to the strong field dependence of the effective mass, in particular above 15~T.

The $g$-factor analysis is performed for the field values and directions, where the splitting of branch $\beta$ is not resolved. The effective mass used for the analysis was measured and the same field interval and is thus an average mass of spin-up and spin-down electrons. The obtained effective $g$ factor is defined  as $g_{\rm eff} = \frac{1}{\mu_0\mu_B}(dE_{F_\uparrow}/dH - dE_{F_\downarrow}/dH)$ and is thus an average effective $g$ factor of both spins. The effective $g$ factor may depend of the spin in the vicinity of the $c$ axis as both spin shows different field dependence of the quantum oscillations frequencies.

 Figure~\ref{g-beta}(b) shows the angular dependence of the $g$-factor $g_\beta,$ in the field interval from 12~T--15~T for angles from [110] to [001] and from [001] to [100]. It depends little on the angular dependence of $m^\star$: an almost similar angular dependence is obtained by taking the angular dependence of the effective mass as obtained in Refs.~\onlinecite{Hassinger2010, Aoki2012} (open circles).  Near to [001] we could not determine the $g$-factor from the spin-splitting zeros due to the non-linear splitting of the $\beta$ branch with field, and the observation of two different frequencies ($\beta_\uparrow$ and $\beta_\downarrow$) for $H \parallel [001]$. As discussed above, the number of spin-splitting zeros for the $\beta$ branch is reduced under magnetic field. This  field dependence is a consequence of the polarization of the small and heavy electron Fermi-surface pockets $\beta$ under magnetic field along the easy magnetization $c$-axis.\cite{Aoki2012, Scheerer2014} Thus we plot in Fig.~\ref{g-beta} the analysis of the $g$-factor for the $\beta$ branch also in the field interval 6~T$ < H < $ 9~T in the angular range from [001] to [100]. Its extrapolation up to [100] gives a very strong $g$-factor variation $g_\beta^{[001]}-g_\beta^{[100]}\approx 2.4$. Between [001] and [110], the oscillation could not be detected in the field interval 6~T--9~T. \deleted{However, the splitting of the $\beta$ oscillation frequency in the interval 12$~$T-15$~$T allows  to deduce the variation of the $g$-factor from the phase of the minority spin pocket by Eq.~(\ref{phaselinearup}). The resulting variation of $g$ is similar to the variation of the $g$-factor between [001] and [100].} The variation of the $g$-factor  for $\beta'$ branch in the field interval 12$~$T-15$~$T with angle from [001] to [100] is also represented in Fig.~\ref{g-beta}. The effective mass for $\beta'$ could be measured only under magnetic field along [100] and we found $m^\star_{\beta'}=20.6~m_0$. This mass is considered as angle independent, too. The angular variation of the $g$-factor for the $\beta'$-branch  appears identical to that of the $\beta$-branch, within the error bars. 
 
 \begin{figure}[!h]
\begin{center}
\includegraphics[width=1\linewidth]{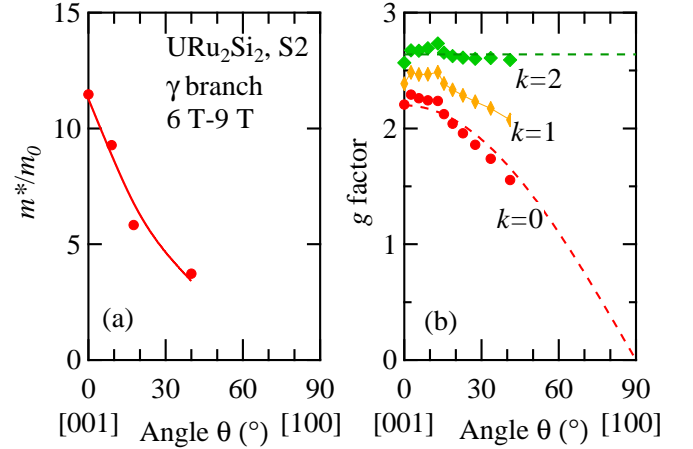}
\caption{(a) Angular dependence of the effective mass of the $\gamma$ branch in the field range 6~T--9~T. (b) Different possibilities for the  angular dependence of the $g$-factor (see text). 
} 
\label{g-gamma}
\end{center}
\end{figure}
 
To analyze the $g$-factor anisotropy of the $\gamma$ branch, the strong anisotropy of its effective mass has to be taken into account, it is shown in Fig.~\ref{g-gamma}(a). The effective mass decreases strongly with angle from $m^\star_\gamma = 11.5 m_0$ at [001] to $m^\star_\gamma = 4.5 m_0$ at 40$^\circ$ to [100]. If the value of $m^\star g/2m_0$ at the first detected amplitude maximum from [001] is $k_\gamma=0$, then the $g$-factor decreases with angle and would reach zero around [100]. The $g$-factor of the $\gamma$ pocket for this scenario is represented in Fig.~\ref{g-gamma}(b) (red circles). In this case, its angular dependence could be fitted by $g_\gamma(\theta)=g_\gamma([001]) \cos(\theta)$ with $g_\gamma([001])=2.2$ corresponding to an Ising behavior of the quasiparticles. However, if we choose $k_\gamma=2$ the occurrence of the spin-splitting zeros can be explained from the anisotropy of the effective mass, and the data can be fitted with a constant $g$-factor $g_\gamma=2.6$. This shows that the $g$-factor determination from the quantum oscillations is generally ambiguous.\deleted{ and special attention has to be taken.}

\subsection{Anisotropy of the $g$-factor}

By quantum oscillation experiments we have been able to investigate the conduction electron $g$-factor of URu$_2$Si$_2$ selectively for different Fermi-surface pockets. For the $\alpha$ Fermi pocket our results are compatible with a rather large $g$-factor ansisotropy. We could show that the angular dependence is not universal between [001] and the basal plane, resulting for the $\alpha$-branch, to an unexplained anisotropy in the basal plane.  
From the present experimental situation, it is not possible to make definite conclusions  on the values of the $g$-factor of the $\alpha$ Fermi surface pocket. The set of values: $g_\alpha^{[001]} \approx 2.5$, $g_\alpha^{[100]} \approx 0$, $g_\alpha^{[110]} \approx 1$ is only a possible solution, \added{under the assumption of a monotonously varying $g$-factor from the $c$ axis to the basal plane} (see Fig.~\ref{g-factor}). \added{However, this would imply a strong anisotropy in the basal plane, which is not observed here. Furthermore, we also did not observe any anisotropy of the upper critical field in the plane what supports a constant $g
$-factor in the basal plane.} \deleted{Thus,} New high-field experiments in the basal plane in a larger field range than studied here, with perfect orientation with respect to the $c$ axis, may resolve directly the observed anisotropy. \deleted{We want to notice that even the rotational symmetry breaking in the basal from four-fold symmetry in the paramagnetic state to a two-fold symmetry in the hidden order state \cite{Okazaki2011, Tabata2014, Kambe2015,Riggs2015, Walstedt2016} \deleted{is still under discussion and} seems experimentally not settled. and demands for a new set of experiments.    }

\added{The $g$-factor for the $\beta$ Fermi-surface pocket is also highly anisotropic. The analysis in the field range from
6~T to 9~T suggests that $g_\beta$ is varying from $g_\beta=0$ in the plane to $g_\beta =2.4$ for $H \parallel [001]$. Interestingly, the determination of the $g$-factor seems dependent on the magnetic field. From the analysis of the spin-splitting zeros in the field range from 12~T to 15~T, a possible solution is a vanishing $g$-factor in the basal plane and $g_\beta= 1.5$ along
the c axis. This field dependence of the measured $g$ factor may be an experimental artefact coming from the field and spin dependence of the effective mass of the $\beta$ branch, which could not be precisely determined in this study and was neglected in the extraction of the $g$ factor. Under these conditions the most reliable value for the $g$ factor of the $\beta$ branch would be the one extracted on the field interval 6~T -- 9~T. In this field range the $g$-factor variation for the $\beta$ branch is  similar to the variation of $g_\alpha$ and one possible solution for $g_\gamma$ in the same plane. We point out that
even the angular dependence of $g$ for the heavy branch $\beta \prime$  show the same anisotropy. Thus we can conclude that
the $g$-factor of all Fermi surfaces show a strong angular dependence. However this variation of $g$ is
slightly bigger to that determined from the weak coupling analysis of the upper critical field $g^{[001]}-g^{[100]}=1.5$. 
}


A relativistic DFT calculation predicted an Ising behavior for the band-like $5f$ electrons in URu$_2$Si$_2$ with magnetic moments along the $c$ axis and no anisotropy in the basal plane.\cite{Oppeneer2011, Werwiniski2014} Here, the 5$f$ electrons are treated as fully itinerant and the calculation is performed for the antiferromagnetic phase which has practically the same Fermi surface than the hidden order state.\cite{Hassinger2010, Elgazzar2009, Oppeneer2010, Ikeda2012} This is justified as the Fermi surfaces for  the localized $5f^2$ or for the localized $5f^3$ uranium configuration are not in correspondence to the experimentally observed ones.\cite{Oppeneer2010} Furthermore, the Fermi surface pockets  obtained in the  itinerant 5$f$  picture are in agreement with all quantum oscillation and ARPES experiments. The Ising anisotropy of the quasiparticles in the DFT calculation is a result of  the peculiar Fermi surface nesting at the hidden order transition and of the strong spin-orbit coupling. All uranium $5f$ states have mainly a total angular momentum $j = 5/2$, and in the paramagnetic state each of the Fermi surface pocket important for the nesting at the hidden order transition have a specific $j_z= \pm 5/2$ or $j_z = \pm 3/2$ character with almost no mixing.  
\cite{Oppeneer2011, Ikeda2012} Due to the doubling of the unit cell,\cite{Hassinger2010, Yoshida2010, Buhot2014}  and concomitant gap opening at the hidden order transition,\cite{Maple1986, Schmidt2010, Aynajian2010} electronic band-structure calculations show that most of the Fermi surface with $j = \pm 5/2$ character is lost and the $\alpha$ and $\beta$ pockets have mainly $j_z = \pm 3/2$ components. Only the pockets at the $M$ point have a $j_z = \pm 1/2$ character.\cite{Ikeda2012}  \added{If the $j_z = \pm 1/2$ component is dominant then $g^{[100]}$ or $g^{[110]}$ will be larger than $g^{[001]}$. }

A different theoretical approach claims that the Ising quasiparticles in URu$_2$Si$_2$ result from the hybridization of the conduction electrons with Ising non-Kramers 5$f^2$ doublet states of the uranium atoms\cite{Chandra2013, Chandra2018} starting from a localized picture of the 5$f$ electrons. However, recent nonresonant inelastic X-ray scattering experiments show that the ground state consists mainly of singlet states in the U$^{4+}$ $5f^2$ configuration.\cite{Sundermann2016} 

The $g$ tensor has never been determined for any heavy fermion system from electronic band-structure calculations. A main difficulty is to know the real crystalline electric ground state of the magnetic ions. 
Furthermore, in URu$_2$Si$_2$, heavy bands are formed due to the strong hybridization of the $s,p,d$ states with the 5$f$ states. Therefore the crystalline field levels are broadened and not clearly observed in spectroscopic experiments. In the localized approach  for a U$^{4+}$ ($5f^2$) configuration, the Land\'{e}'s $g_J$-factor in an intermediate coupling regime is $g_J=0.824$ and for a U$^{3+}$ ($5f^3$) configuration $g_J=0.744$.\cite{Amoretti1984} First-principles dynamical mean field  calculations concluded that for URu$_2$Si$_2$, the 5$f^2$ configuration has the dominant weight.\cite{Haule2009} The multiplet of the 5$f^2$ has a total angular momentum $J = 4$ and splits into five singlets and two doublets. The doublets are linear combinations of the $|J_z=\pm 3\rangle$ and $|J_z=\pm 1\rangle$.\cite{Ohkawa1999} The lowest doublet is $|\!\!\uparrow \rangle = \cos\theta |J_z=  3\rangle + \sin\theta |J_z= -1\rangle $  
and $|\!\!\downarrow \rangle = \cos\theta |J_z= -3\rangle + \sin\theta |J_z= 1\rangle $, with $\theta$ being the angle between the $c$ axis and the basal plane.
In this case the $g$-factors are anisotropic and $g_z = g_J(3\cos^2\theta - \sin^2\theta)$ and in the basal plane $g_x = g_y = 0$. However, this $g$-factor in the fully localized picture has never been observed. 

 This localized approach has been discussed in Ref.~\onlinecite{Altarawneh2012} and the authors have fitted the $g$-factor anisotropy of the $\alpha$ pocket in the angular range from [001] to [100] and get $\cos\theta = 0.8$. 
As mentioned above, the $g$-factor determined by quantum oscillations is Fermi surface selective, and results from an average of the $g$-factor of electrons on the orbit perpendicular to the applied magnetic field. We have shown that the $g$-factor for all detected Fermi surfaces are consistent with a strong $g$-factor anisotropy.   

\subsection{Comparison to other heavy-fermion system}

The determination of the $g$-factor in heavy-fermion systems is rare. A standard method to determine the $g$-factor in magnetic insulators is electron spin resonance (ESR). However, a narrow ESR line in Kondo lattices have been only reported in some Yb- or Ce-based compounds which show very strong ferromagnetic fluctuations such as e.g.~YbRh$_2$Si$_2$ or CeRuPO.\cite{Sichelschmidt2003, Krellner2008} Several theories are devoted to explain the line-width narrowing in these systems starting from a localized or an itinerant model approach.\cite{Kochelaev2009, Woelfle2009, Schlottmann2009}. In these systems, the large anisotropy of the $g$-factor reflects the local anisotropy in the intersite correlations. In YbRh$_2$Si$_2$ the anisotropy of the local $g_f$ factor of the Yb ion is about a factor 20, $g_{f\perp} =3.6$ and $g_{f\parallel}=0.17$ and refects the large anisotropy of the susceptibility.
 
Spin-splitting zeros have been used to determine the angular dependence of the $g$-factor in simple metals such as gold or copper (see e.g.~Ref.~\onlinecite{Randles1972, Higgins1980}). \footnote{While in Cu the $g$-factor is isotrope, a spin-splitting zero appears near 13 deg.~from [111] due to the anisotropy of the effective mass. In the noble metals like Au, the $g$-factor is anisotropic. \cite{Randles1972, Higgins1980}} 
Whereas quantum oscillations are studied for almost every heavy-fermion system which could be grown in sufficiently high quality, the observation of spin-splitting zeros and so the determination of the $g$-factor is very rare. Especially in systems showing strong Ising-type anisotropy, it has never been observed. In CeRu$_2$Si$_2$, the best studied example, it has not been observed although the Fermi surface has been determined in great detail by quantum oscillation experiments (for a  review see Ref.~\onlinecite{HAoki2014}). This may be due to topology difference of the \added{spin-up and spin-down} Fermi surfaces. In URu$_2$Si$_2$ only small closed Fermi surface pockets exist in the hidden order state, whereas in CeRu$_2$Si$_2$ large pockets are detected, and also open Fermi surfaces exist. 
 
Spin-splitting zeros have been observed in the cubic CeIn$_3$, which orders antiferromagnetically below 10~K. One of its dHvA branches, named $d$, which corresponds to a closed spherical Fermi surface centered at the $\Gamma$ point in the Brillouin zone, has a highly anisotropic cyclotron effective mass. While the effective mass is about 2-3$m_0$ for $H \parallel [100]$, it reaches 12--16~$m_0$ for $H \parallel [110]$.  In CeIn$_3$ the determination of the $g$-factor from the spin-splitting zeros of the dHvA oscillations has not been unambiguous, because  of the integer $k$ for $g (m^\star/m_0) = 2(n+k)-1$ .\cite{Settai1995} The effective mass is usually isotropic, if the topology of the Fermi surface is spherical in a highly symmetric crystal structure such as a cubic system. In CeIn$_3$, this anisotropic effective mass on the spherical Fermi surface is probably due to the consequence of strong electron correlations with anisotropic 4$f$-contribution on the Fermi surface leading to hot spots at the antiferromagnetic wave vector. \deleted{In YRu$_2$Si$_2$ , the $g$-factor was determined for branch $\beta$, which corresponds to the hole Fermi surface centered at $Z$ point.\cite{Ikezawa1995} The topology of the Fermi surface is simply ellipsoidal, and the angular dependence of the effective mass was assumed to follow the so-called effective mass model, using the effective masses, 0.33~$m_0$ and 0.70~$m_0$ detected experimentally for $H \parallel [001]$ and [100], respectively. A spin-splitting zero was observed only at 30$^\circ$ tilted from [001] to [100]. From these simple experimental results, the $g$-factor was derived to be $g=6.2$, and is constant from [001] to [100].}

\subsection{Relation between $g$-factor anisotropy and hidden order and superconductivity}
\label{discussion_superconductivity}

As pointed out in Ref.~\onlinecite{Mineev2015}, the strong uniaxial $g$-factor anisotropy is also compatible with a non-conventional commensurate charge density wave. Recently a chirality-density wave has been proposed as order parameter of the hidden-order state from Raman-scattering experiments, where a particular inelastic excitation with $A_{2g}$ symmetry has been observed.\cite{Buhot2014,Kung2015} The proposed density wave is in agreement with the previously determined folding of the Brillouin zone along the $c$ axis at the hidden order transition and confirms the change from a body-centered-tetragonal to a simple-tetragonal electronic structure. For commensurate antiferromagnetically ordered systems, it appears that due to the anisotropic spin-orbit character of the Zeeman coupling, the transverse component of the $g$ tensor shows a significant momentum dependence: it vanishes in the plane perpendicular to the direction of the staggered magnetization due to a conspiracy of the crystal symmetry with that of the antiferromagnetic order.\cite{Brazovskii1989, Ramazashvili2008,  Ramazashvili2009} 
If such a scenario is valid for the hidden order state with a characteristic ordering vector $Q_{HO} = (001)$, the appearance of the spin-splitting zeros would not be due to a local property of the U - ion but to a collective ordering in the hidden order state. 

This would also explain why almost the same anisotropy of the electronic $g$-factor is observed on the different Fermi-surface pockets. The remaining differences are due to differences in the effective mass and to details in the band structure, which results in a momentum dependent spin-orbit coupling.

\begin{table}[h]
\begin{ruledtabular}
\caption{\label{table}Fermi velocities $v_F = \frac{\hbar}{m^\star}\sqrt{\frac{2\pi F}{\Phi_0}}$ for different Fermi surface pockets in URu$_2$Si$_2$.  $F$ and $m^\star$ are the oscillation frequency and the effective mass of the corresponding Fermi surface pocket. $v_F^{[001]}$, $v_F^{[100]}$ (m/s), and $v_F^{[110]}$ (m/s) give the Fermi velocities for field applied $H$ along [001], [100], and [110], respectively.  }
\begin{tabular}{ccccc}
		      & $v_F^{[001]}$ (m/s)	& $v_F^{[100]}$ (m/s)	& $v_F^{[110]}$ (m/s) &	\\
		      \hline
	$\alpha$	& 16340 				& 24062				& 20860 & (this work) \\
	$\beta$	& 6625 				& 5157				& 6070  & (this work) \\
	$\gamma$& 10760				& 8000				& 9450 & (Ref.~\onlinecite{Aoki2012}) \\
	$\eta$	& 4000 				& 					& 		& (Ref.~\onlinecite{Hassinger2010}) 
\end{tabular}
\end{ruledtabular}
\end{table}
\added{
Finally, we want to compare the $g$ factor anisotropy determined from quantum oscillations with that deduced from the anisotropy of the upper critical field. The $g$-factor determined from the paramagnetic limitation of the upper critical field $H_{c2}$ gives an average over all the Fermi-surface pockets contributing to the superconducting state. 
Near $T_c$, the observed initial slope of the upper critical field near $T_{sc}$ for a clean superconductor, $H_{c2}'\propto \frac{T_{sc}}{v_F^2}$, where $v_F$ is an average Fermi velocity perpendicular to the applied field. The Fermi velocities can be determined from the quantum oscillation experiments and are given in Tab.~\ref{table} for different Fermi surface pockets.
It is obvious that the strong anisotropic initial slope of the upper critical field cannot be explained by the anisotropy of the observed Fermi velocities. 
Indeed, to explain the factor 2.2 of anisotropy between $H’_{c2}$ for $H \parallel c$ or $H_\perp$, a factor ~1.5 is required on the corresponding Fermi velocities.
For the$\beta$ branch, which has the smallest Fermi velocity observed, the values of $v_F \perp H$ are of the right order to explain the measured value of $H’_{c2} \approx (-11 {\rm T/K})$, applying formulas for a spherical Fermi surface and $s$-wave superconductivity ($v_F = 6050 {\rm m/s}$). But then, a value of 8900 m/s would be required for $H \parallel c$, much larger than the actual value. This points out the difficulty of precise quantitative comparisons between measured normal state properties and $H_{c2}$ measurements: already for $s$-wave superconductors, it is known that the average $v_F$ determining $H_{c2}$ involves an average over all Fermi sheets weighted by the pairing potential \cite{Langmann1992, Kita2004}. In case of the proposed $d$-wave pairing \cite{Kasahara2007, Kittaka2016}, the strong gap anisotropy may play a dominant role in the determination of the orbital anisotropy of $H_{c2}$. However, numerical calculations are required, as well as a complete determination of the Fermi surface of URu$_2$Si$_2$: the heaviest mass -- the anisotropic electron Fermi surface centered at the $M$ point of the simple tetragonal Brillouin zone\cite{Ikeda2012} --  and so possibly the dominant FS sheets for the control of $H_{c2}$ are still not detected in the quantum oscillations.\footnote{This heavy orbit may be observed by cyclotron resonance experiments.\cite{Tonegawa2012, Tonegawa2013}}
}


A next important step in understanding the Fermi surface and its feedback on the hidden order would  be to determine completely the Fermi surface in the high pressure antiferromagnetic state. It is known from SdH experiments that the quantum oscillation frequencies and the effective masses of the main Fermi-surface branches evolve smoothly from the hidden order phase at low pressure to the antiferromagnetic state above 1~GPa.\cite{Nakashima2003, Hassinger2010} A detailed study of the angular dependence under high pressure will show whether the observed anisotropy of the $g$-factor is a particular characteristic of the hidden order, or not.

\section{Conclusion}

We have determined selectively the electronic $g$-factor and its anisotropy for the $\alpha$, $\beta$, and $\gamma$ Fermi surface pockets of URu$_2$Si$_2$ between [001] and the basal plane. 
For all detected Fermi surface pockets, our results are consistent with a strongly anisotropic $g$-factor. 
\deleted{For  the $\alpha$ pocket we observed a remaining anisotropy of the $g$-factor in the basal plane with a finite $g$ value for a magnetic field applied along the [110] direction, while it can be zero along [100].} 
For the $\beta$ and $\gamma$ branches, possible solutions exist with vanishing in plane $g$-factor. 
\added{For the $\alpha$ branch, we observed different numbers of spin-splitting zeros in the (010) and (110) planes, which indicate either a non-monotonous variation of the $g$ factor in one of these planes or an additional anisotropy in the basal plane.  Future experiments in high magnetic fields have to be performed to clarify the $g$-factor anisotropy of the $\alpha$ branch.} \deleted{These results, compared with the anisotropy of the $g$-factor determined from the upper critical field, show the importance of the heavy branches for the superconducting pairing. } \added{The determined anisotropy of the $g$-factor by quantum oscillations is in good agreement with that from the superconducting upper critical field. However, the anisotropy of the initial slope of the upper critical field cannot be explained simply by the observed Fermi surface pockets. An anisotropic heavy Fermi surface pocket still has not been detected in quantum oscillations. 
The reported determination of the anisotropy of the $g$-factor by quantum oscillations} is an important reference for other heavy fermion systems, showing that itinerant quasiparticles in a metal can have a very strongly anisotropic $g$-factor (Ising-like). Moreover, we hope that our results will stimulate calculations of the $g$-factor from the  electronic band structure.

\section{Acknowledgements}
We thank H.~Harima, H.~Ikeda, G.~Zwicknagl, V.P.~Mineev, J.P.~Sanchez, H.A.~Krug von Nidda, E.~Hassinger, \added{A.~Pourret}, P.~Chandra, and P.~Oppeneer for valuable and fruitful discussions. Furthermore, we are grateful to H.~Harima for critical reading of the manuscript. This work has been supported by ERC (NewHeavyFermion), KAKENHI (JP15H05882, JP15H05884, JP15K21732, JP16H04006, JP15H05745).

\bibliography{uru2si2}

\end{document}